\documentclass{elsarticle}
\usepackage{graphicx}
\usepackage{amsmath}
\usepackage{amssymb}
\usepackage{hyperref}
\usepackage{color}
\usepackage[dvipsnames]{xcolor}
\newcommand{\rev}[1]{#1}
\newcommand{\revtwo}[1]{#1}
\usepackage{microtype}
\usepackage{siunitx}
\usepackage{subcaption}
\setcitestyle{square, comma, numbers, sort&compress, super}
\usepackage[version=3]{mhchem}
\usepackage{geometry}
\usepackage{balance}
\usepackage{mathptmx}
\usepackage{sectsty}

\usepackage{lastpage}
\usepackage[format=plain,justification=justified,singlelinecheck=false,
font={stretch=1.125,small,sf},labelfont=bf,labelsep=space]{caption}
\usepackage{float}
\usepackage{fancyhdr}
\usepackage{fnpos}
\usepackage[english]{babel}
\addto{\captionsenglish}{%
  
}
\usepackage{array}
\usepackage{droidsans}
\usepackage{charter}
\usepackage[T1]{fontenc}
\usepackage{anyfontsize}
\usepackage{mathtools}
\usepackage{amsfonts}
\usepackage{listings}
\usepackage{inconsolata}

\usepackage{natbib}
\usepackage{orcidlink}
\usepackage{afterpage}
\newcommand{\tags}[1]{%
    \vspace{-1em}%
    \begin{center}
        \textit{Tags: #1}
    \end{center}
}

\usepackage{titlesec}
\journal{Computer Physics Communications}

\titleformat{\section}[block]{\normalfont\LARGE\bfseries}{\hspace{0em}}{0em}{}
\titleformat{\subsection}[block]{\normalfont\Large\bfseries}{\hspace{0em}}{0em}{}
\titleformat{\subsubsection}[block]{\normalfont\large\bfseries}{\hspace{0em}}{0em}{}

\lstdefinelanguage{json}{
  basicstyle=\ttfamily\small,
  numbers=left,
  numberstyle=\scriptsize,
  stepnumber=1,
  showstringspaces=false,
  breaklines=true,
  frame=single,
  morestring=[b]",
  stringstyle=\color{teal},
  morecomment=[l]{//},
  commentstyle=\color{gray},
}

\begin{document}

\begin{frontmatter}
\title{
$\kappa$ALDo 2.0: Scalable thermal transport from first principles and machine learning potentials
}

\author{Giuseppe Barbalinardo\orcidlink{0000-0002-7829-4403}\fnref{eqc}$^{a}$}
\ead{gbarbalinardo@ucdavis.edu}
\author{Zekun Chen\orcidlink{0000-0002-4183-2941}\fnref{eqc}$^{a}$}
\author{Dylan Folkner\orcidlink{0000-0002-8715-7048}$^{a}$}
\author{Bohan Li\orcidlink{0009-0000-9946-630X}$^{a, c}$}
\author{Nicholas Lundgren\orcidlink{0000-0003-1914-0954}$^{a}$}
\author{Nathaniel Troup\orcidlink{0009-0002-9789-2238}$^{a}$}
\author{Alfredo Fiorentino\orcidlink{0000-0002-3048-5534}$^{b,d}$}
\author{Davide Donadio\orcidlink{0000-0002-2150-4182}$^{a}$}
\affiliation{Department of Chemistry, University of California, Davis, Davis, CA, 95616, USA}
\affiliation{PSI Center for Scientific Computing, Theory and Data, 5232 Villigen PSI, Switzerland}
\affiliation{Massachusetts Institute of Technology, Cambridge, MA, USA}
\affiliation{National Centre for Computational Design and Discovery of Novel Materials (MARVEL), 5232 Villigen PSI, Switzerland}
\fntext[eqc]{These authors contributed equally to this work.}
\ead{ddonadio@ucdavis.edu}
\date{\today}

\begin{abstract}
We introduce $\kappa$ALDo~2.0, an open-source Python package for computing vibrational, elastic, and thermal transport properties of crystalline and disordered solids from first principles and machine-learned interatomic potentials. Building on the anharmonic lattice dynamics (ALD) framework, $\kappa$ALDo~2.0 provides efficient CPU and GPU-accelerated implementations of the Boltzmann transport equation (BTE) for crystals and the quasi-harmonic Green-Kubo (QHGK) method. The QHGK formalism extends thermal transport predictions beyond translationally-invariant crystals to materials lacking long-range order, including glasses, alloys, and complex nanostructures. $\kappa$ALDo~2.0 introduces native integration with modern machine-learned potentials (MLPs), enabling thermal transport workflows that combine the accuracy of first-principles methods with the scalability of classical force fields. It also features comprehensive support for temperature-dependent effective potentials (TDEP) workflows, flexible storage backends for large-scale calculations, and advanced quantification of anharmonicity.

The software seamlessly interfaces with electronic structure codes (Quantum ESPRESSO, VASP), molecular dynamics packages (LAMMPS), and state-of-the-art MLPs (ACE, NEP, MACE, MatterSim, Orb), enabling thermal transport studies from \SI{0}{\kelvin} to finite temperatures. $\kappa$ALDo~2.0 implements multiple BTE solution strategies (relaxation time approximation, self-consistent iteration, full matrix inversion, and eigendecomposition) and supports essential physical corrections, including isotopic scattering and non-analytical terms for polar materials. A modular Python architecture with lazy evaluation and multiple storage formats (formatted text, NumPy, HDF5) enables simulations of systems containing more than 10,000 atoms.

This paper describes the theoretical framework, implementation details, software architecture, and validation examples demonstrating $\kappa$ALDo~2.0's capabilities for studying complex materials, including halide perovskites with strong anharmonicity and polar oxides requiring long-range electrostatic corrections.
\end{abstract}
\end{frontmatter}

\tags{Python, Materials Science, Computational Physics, Lattice Dynamics, Thermal Conductivity, Boltzmann Transport Equation, Machine Learning Potentials, Green-Kubo, Phonon Scattering, GPU Acceleration, TDEP, Anharmonic Effects}

\noindent\textbf{Program Title:} $\kappa$ALDo 2.0 \\
\textbf{CPC Library link to program files:} \url{https://doi.org/10.17632/t3f42nv4fw.1} \\
\textbf{Developer's repository:} \url{https://github.com/nanotheorygroup/kaldo} \\
\textbf{Examples repository:} \url{https://github.com/nanotheorygroup/kaldo-examples} \\
\textbf{Licensing provisions:} BSD 3-Clause License \\
\textbf{Programming language:} Python 3.10+ \\
\textbf{External libraries:} NumPy~\cite{harris2020array}, SciPy~\cite{virtanen2020scipy}, TensorFlow~\cite{tensorflow2015}, ASE~\cite{larsen_atomic_2017}, sparse, opt\_einsum~\cite{smith2018opt}, h5py~\cite{collette2014h5py}, seekpath~\cite{hinuma2017seekpath} \\

\noindent\textbf{Nature of problem:}
Accurate prediction of lattice heat transport in solids requires the calculation of anharmonic interatomic force constants, phonon scattering processes, and solution of the Boltzmann transport equation (BTE) for crystals or alternative formalisms for disordered materials where BTE assumptions break down. A modern software to compute lattice thermal conductivity needs: (i) flexibility to handle both crystalline and disordered materials within a unified framework, (ii) support for finite-temperature renormalization effects critical in phase-changing materials, (iii) integration with modern machine-learned potentials, or (iv) scalability to systems with thousands of atoms. A unified platform is needed that seamlessly extends the calculation of lattice thermal conductivity from crystals, including strongly anharmonic crystals, to amorphous solids while combining first-principles accuracy with computational efficiency.

\noindent\textbf{Solution method:}
$\kappa$ALDo~2.0 implements anharmonic lattice dynamics with modular support for multiple force constant generation methods (finite differences, perturbation theory, fitting from MD trajectories) and thermal conductivity solvers: BTE (RTA, self-consistent, full matrix inversion, eigendecomposition) for crystalline materials, and quasi-harmonic Green-Kubo (QHGK) for strongly anharmonic and disordered systems. The QHGK implementation extends lattice dynamics predictions to materials lacking translational symmetry, bridging the gap between traditional BTE and molecular dynamics. The code uses sparse tensor representations and GPU acceleration via TensorFlow for the computationally intensive projection of third-order force constants onto phonon eigenstates. Multiple storage backends enable memory-efficient caching of intermediate results. Integration with ASE and various native interfaces (LAMMPS, Quantum-Espresso, etc.) provides access to diverse force calculators, including modern MLPs.

\noindent\textbf{Additional comments:}
$\kappa$ALDo 2.0 is particularly suited for: (i) materials with strong temperature-dependent anharmonicity (e.g., halide perovskites, materials near phase transitions), (ii) studies requiring systematic comparison of multiple thermal conductivity solution methods, (iii) large-scale materials screening workflows, and (iv) development and testing of new theoretical methods in phonon transport. The software includes extensive test coverage, Docker deployment, Google Colab tutorials, and auto-generated API documentation. No special hardware is required, though GPU access significantly accelerates calculations for systems with unit cells containing over 100 atoms.

\section{Introduction}

Thermal management has emerged as a critical challenge in modern technology, from microelectronics miniaturization to thermoelectric energy conversion and thermal barrier coatings. As device dimensions approach the nanoscale, classical continuum approaches to heat transport break down, necessitating atomistic descriptions of phonon-mediated thermal conductivity~\cite{cahill2003nanoscale,cahill_nanoscale_2014}. Understanding and predicting thermal transport at the atomistic level is essential for materials design in applications ranging from heat dissipation in high-power electronics to efficiency optimization in energy conversion devices~\cite{shi_evaluating_2015}.

The past decade has witnessed remarkable progress in computational materials science driven by three converging developments. First, large-scale materials databases~\cite{curtarolo2012aflow, jain_commentary_2013, kirklin2015open} have enabled systematic screening of compounds across chemical space, identifying promising candidates for targeted properties. 
Second, advances in high-performance computing and algorithmic improvements have extended the reach of \emph{ab initio} methods to increasingly complex systems~\cite{carnimeo2023quantum}.
Third, machine learning has revolutionized interatomic potential development, initially with specialized models that achieve near-DFT accuracy at computational scaling and overall cost comparable to empirical potentials~\cite{behler_generalized_2007,bartok_gaussian_2010,shapeev_moment_2016,thompson_spectral_2015,zhang_deep_2018,drautz_atomic_2019,fan_neuroevolution_2021, batzner_e3-equivariant_2022}, 
and, more recently, with foundation models that cover a large portion of the periodic table~\cite{chen_universal_2022,yang_mattersim_2024,duschatko_orb_2024,batatia2025foundation}. 


In semiconductors and insulators, thermal transport is governed by phonons, quantized lattice vibrations arising from collective atomic motion. Molecular dynamics (MD) provides a direct route to compute thermal conductivity, either through the equilibrium Green-Kubo formalism (EMD)~\cite{green_markoff_1954,kubo_statistical-mechanical_1957,zwanzig_time-correlation_1965,ladd_lattice_1986} {or via non-equilibrium approaches (NEMD) that impose a temperature gradient across the simulation cell~\cite{muller-plathe_simple_1997,schelling_comparison_2002}. A key strength of MD is that it captures the anharmonicity of the potential energy surface to all orders, making no perturbative truncation. Moreover, MD is straightforwardly applicable to complex systems such as nanostructured, defective, and disordered materials, where the lack of translational symmetry makes reciprocal-space methods impractical. Both EMD}~\cite{sosso_thermal_2012,mangold_transferability_2020,langer_heat_2023,dong_molecular_2024} {and NEMD can be combined with empirical, machine-learned, or} first-principles~\cite{marcolongo_microscopic_2016} force evaluations. However, {MD-based methods face} fundamental and practical limitations: (i) {classical MD} cannot capture quantum phonon populations, (ii) {achieving size and time convergence may involve} high computational cost, particularly for NEMD where finite-size extrapolation is required~\cite{schelling_comparison_2002}, and (iii) decomposing the total conductivity into mode-specific contributions is challenging, thus limiting physical insight.

Anharmonic lattice dynamics (ALD) offers a complementary approach that addresses these limitations~\cite{ziman2001electrons, srivastava2022physics}. By computing phonon frequencies, lifetimes, and transport properties from interatomic force constants (IFCs), ALD achieves favorable scaling: computational cost grows with the number of phonon modes ($3N$, where $N$ is the number of atoms) rather than MD timesteps. The Boltzmann transport equation (BTE) framework~\cite{ziman2001electrons} enables systematic treatment of phonon-phonon scattering, isotopic disorder, and boundary effects, while providing mode-resolved analysis unavailable from EMD.

Recent theoretical advances have broadened the applicability of ALD. Unified theories, such as the Wigner approach and the quasi-harmonic Green-Kubo (QHGK) method~\cite{simoncelli_unified_2019,isaeva_modeling_2019}, have extended lattice dynamics to strongly anharmonic crystals and disordered materials, including glasses, alloys, and partially disordered nanostructures, for which interband energy transfer provides a substantial contribution to thermal transport~\cite{neogi_anisotropic_2020,lundgren2021mode,simoncelli2022wigner,zhang2022coherence, fiorentino2025effects}\rev{\cite{simoncelli_thermal_2023}}.
Temperature-dependent effective potentials (TDEP)~\cite{hellman_temperature-dependent_2013,castellano2023mode,knoop_tdeptemperature_2024} enable treatment of strong anharmonicity, nuclear quantum effects, and temperature-induced renormalization of both harmonic and anharmonic vibrational properties~\cite{folkner2024elastic}. 

Whereas most of the codes available to perform ALD calculations of thermal transport (ShengBTE\cite{li2014shengbte}, Phono3py~\cite{togo2023first}, Phoebe~\cite{cepellotti_phoebe_2022}, ALAMODE~\cite{tadano_anharmonic_2014}, THERMACOND~\cite{nayeb_sadeghi_thermacond_2025}, and Thermal2~\cite{paulatto2013anharmonic,fugallo_ab_2013}) are designed to target crystalline systems by first principles, 
$\kappa$ALDo 2.0 is created to address complex materials and acts as a native Python bridge among modern MLPs, TDEP, and advanced transport formalisms (BTE and QHGK). It unifies BTE and QHGK methods with seamless integration to modern force calculators (DFT codes, LAMMPS, MLPs) and GPU acceleration. Version 2.0 features: (i) multiple BTE solvers (RTA, self-consistent, full inversion, eigendecomposition), (ii) comprehensive physical corrections (isotopes, non-analytical terms, finite-size effects), (iii) modular storage backends enabling \revtwo{calculations on systems of over 10,000 atoms}, (iv) extensive documentation and containerized deployment, \revtwo{and (v) crystal-symmetry reduction of the Brillouin-zone sampling together with shared-memory parallelization of the force-constant and $q$-point calculations}. \revtwo{These capabilities are summarized in Section~\ref{sec:overview}.}

This paper is organized as follows. Section~\ref{sec:theory} presents the theoretical foundations and computational algorithms. Section~\ref{sec:features} describes advanced physical modeling capabilities and software infrastructure. Section~\ref{sec:applications} demonstrates $\kappa$ALDo~2.0's capabilities through applications to \ce{CsPbBr3} halide perovskite and \ce{MgO} oxide. Section~\ref{sec:conclusion} summarizes key features and future directions.

\section{Theory and Implementation}
\label{sec:theory}

\begin{figure*}[h!]
    \centering
    \includegraphics[width=0.70\textwidth]{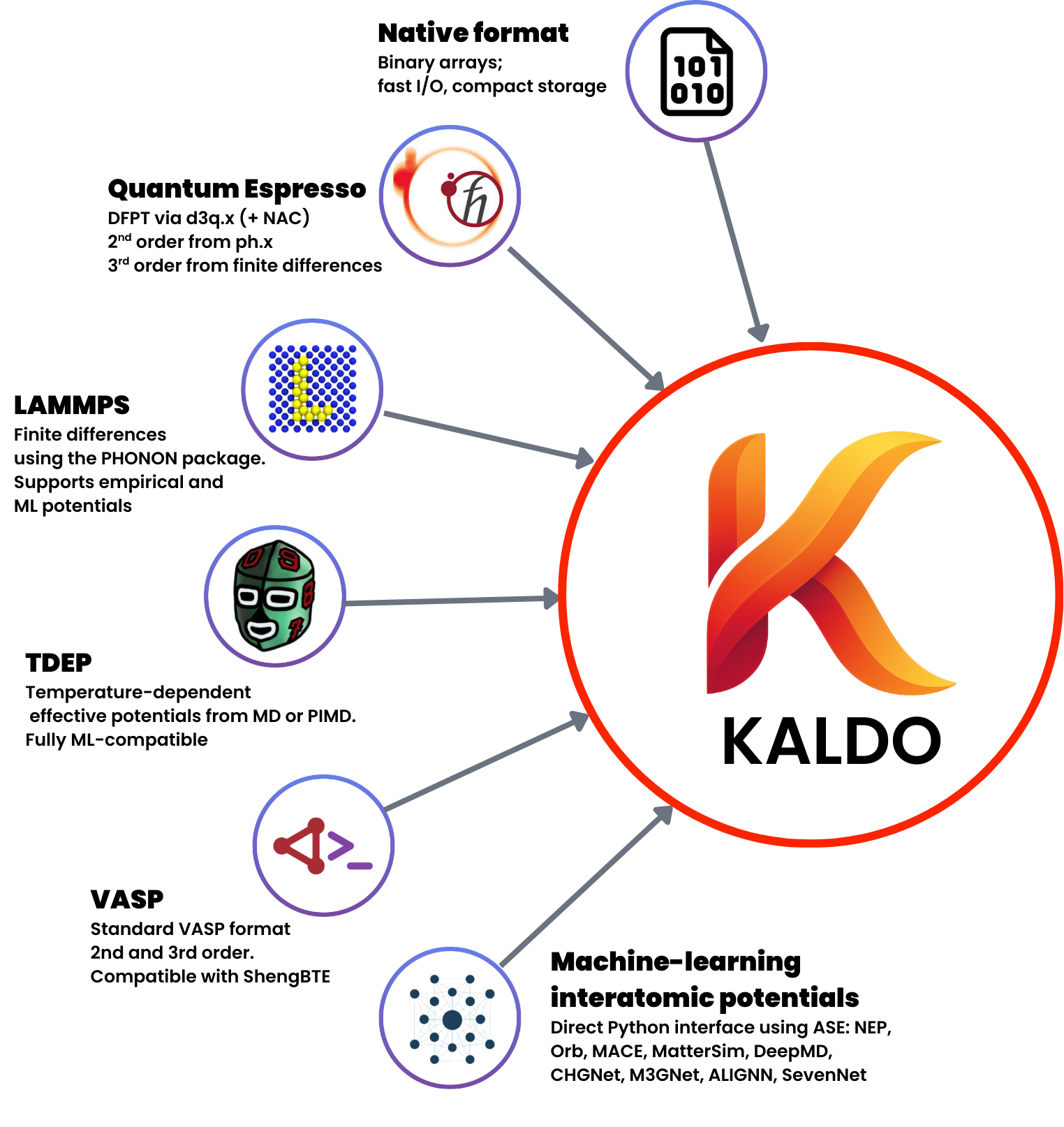}
    \caption{$\kappa$ALDo~2.0 object-oriented class structure. \textbf{ForceConstants} manages second- and third-order IFCs with methods for importing from external codes (\texttt{from\_folder()}) or direct calculation via ASE (\texttt{from\_ase()}). \textbf{Phonons} computes vibrational properties including frequencies, velocities, scattering rates, heat capacities, and populations. \textbf{Conductivity} solves for thermal transport via BTE (RTA, self-consistent, inverse, eigendecomposition) or QHGK methods, with outputs including thermal conductivity tensors, mean free paths, and diffusivities. All classes inherit from \texttt{Storable}, implementing lazy evaluation and flexible storage backends.}
    \label{fig:kaldo-classes}
\end{figure*}

$\kappa$ALDo~2.0 implements a hierarchical workflow for thermal transport calculations, progressing from atomic structure and interatomic force constants through phonon properties to thermal conductivity. This modular design mirrors the underlying physics while enabling flexible, efficient computation across diverse material systems. The three core components, \texttt{ForceConstants}, \texttt{Phonons}, and \texttt{Conductivity} classes, form the foundation of the software architecture (Fig.~\ref{fig:kaldo-classes}).

\subsection{Computational Workflow}

\begin{figure*}[h!]
    \centering
    \includegraphics[width=0.45\textwidth]{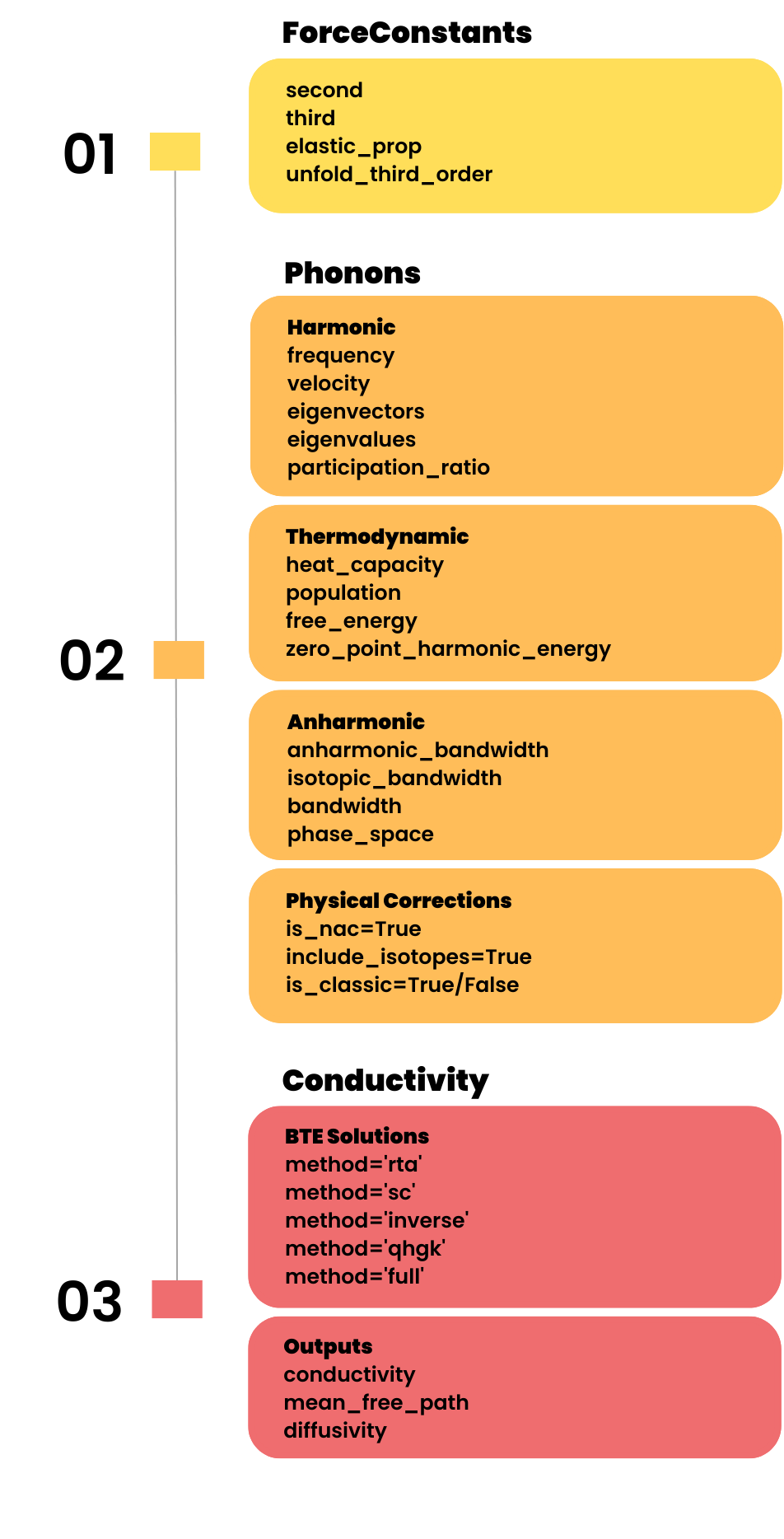}
    \caption{$\kappa$ALDo~2.0 input/output ecosystem. The software interfaces with diverse force constant sources, including \emph{ab initio} codes (Quantum ESPRESSO, VASP), molecular dynamics packages (LAMMPS), TDEP, machine-learned potentials via ASE (NEP, MACE/MatterSim, Orb), and external phonon codes (ShengBTE, phono3py, HiPhive). The workflow progresses from force constant import/generation to phonon property calculation to thermal conductivity evaluation via BTE or QHGK methods.}
    \label{fig:kaldo-inputs}
\end{figure*}

Thermal transport calculations in $\kappa$ALDo~2.0 proceed through three main stages:

\textbf{Stage 1: Force Constant Generation or Import.} Second and third-order interatomic force constants (IFCs) are computed or imported. These tensors, encoding harmonic and anharmonic interatomic interactions, serve as the input for all subsequent analysis. Methods include: (i) finite-difference calculations with user-specified atomic displacements, (ii) direct import from density functional perturbation theory (DFPT) calculations, (iii) extraction from molecular dynamics trajectories via TDEP, or (iv) loading from external codes in various file formats (Fig.~\ref{fig:kaldo-inputs}).

\textbf{Stage 2: Phonon Property Computation.} The dynamical matrix constructed from second-order IFCs is diagonalized to obtain phonon frequencies, eigenvectors, and group velocities. Third-order IFCs are projected onto phonon eigenstates to compute scattering phase space and rates. This computationally intensive step uses sparse tensor operations and optional GPU acceleration. Additional properties, including mode-resolved heat capacity, populations (Bose-Einstein or classical), and optional isotopic scattering rates, are calculated.

\textbf{Stage 3: Thermal Conductivity Calculation.} The BTE or QHGK formalism is employed using phonon properties from Stage 2. Multiple solution methods are available depending on accuracy requirements, memory constraints, and material characteristics. Results include mode-resolved thermal conductivity tensors, mean free paths, and optional diffusivities.

\subsection{Interatomic Force Constants}

The first step of the ALD computation of thermal transport is the calculation of the harmonic and anharmonic force constants. In the standard lattice dynamics approach, these are the second-order, $K^{(2)}=\{\phi^{\prime\prime}_{i\alpha i'\alpha '}\}$, and higher-order derivatives of the potential energy surface with respect to atomic displacements. Anharmonic force constants are usually limited to third derivatives, $K^{(3)}=\{\phi^{\prime\prime\prime}_{i\alpha i'\alpha 'i''\alpha ''}\}$, corresponding to three-phonon processes, although four-phonon processes are critical to capture the correct $\kappa$ of high-conductivity materials, such as diamond, graphene, and BAs~\cite{feng2017four,feng_four-phonon_2018}.
In finite temperature approaches, e.g., TDEP, $K^{(2)}$ and $K^{(3)}$ , are numerical estimates of the derivatives of the Gibbs free energy~\cite{castellano2023mode,folkner2024elastic}.
Either way, $K^{(2)}$ and $K^{(3)}$ are second-order and third-order tensors. 

The mass-scaled second-order IFCs define the dynamical matrix elements:
\begin{equation}
D_{i\alpha i'\beta} = \frac{K^{(2)}_{i\alpha i'\beta}}{\sqrt{m_i m_{i'}}},
\label{eq:dynmat}
\end{equation}
where $m_i$ is the mass of atom $i$. Phonon frequencies $\omega_\mu$ and polarization vectors $\eta_{i\alpha\mu}$ (mode index $\mu$) satisfy the eigenvalue equation:
\begin{equation}
\sum_{i'\beta} D_{i\alpha i'\beta} \eta_{i'\beta\mu} = \omega_\mu^2 \eta_{i\alpha\mu}.
\label{eq:eigenvalue}
\end{equation}

For periodic crystals, translational symmetry enables reciprocal-space formulation. The dynamical matrix at wavevector $\mathbf{q}_k$ is:
\begin{equation}
D_{i\alpha k i'\beta} = \sum_{\mathbf{R}_l} e^{-i \mathbf{q}_k \cdot \mathbf{R}_l} D_{i\alpha li'\beta},
\label{eq:dynmat_reciprocal}
\end{equation}
where $\mathbf{R}_l$ indexes unit cell replicas. The eigenvalue problem becomes:
\begin{equation}
\sum_{i'\beta} D_{i \alpha k i' \beta} \eta_{i' \beta k s} = \omega_{k s}^{2} \eta_{i \alpha k s},
\label{eq:eigenvalue_crystal}
\end{equation}
with band index $s$ and wavevector index $k$. The composite index $\mu = (k,s)$ labels individual phonon modes. The acoustic sum rule, ensuring zero frequency for the three acoustic modes at $\Gamma$ ($\mathbf{q}=0$), can be enforced via the \texttt{is\_acoustic\_sum} flag.

\subsubsection{Implementation and Interfaces}

The \texttt{ForceConstants} class provides a unified interface for importing and manipulating IFC files. The class method \texttt{from\_folder()} supports multiple input formats listed in Table~\ref{tab:formats}. Additionally, the \texttt{from\_ase()} method enables direct IFCs calculations from any ASE-compatible calculator, providing access to all modern machine-learned potentials (NEP, MACE, MatterSim, Orb, etc.) as well as classical force fields and DFT codes. This flexible design enables seamless integration with diverse computational workflows.

\begin{table}[h!]
\centering
\caption{Supported force constant file formats in $\kappa$ALDo~2.0, enabling integration with diverse computational workflows.}
\label{tab:formats}
\small
\begin{minipage}{\textwidth}
\centering
\begin{tabular}{lllll}
\hline
\textbf{Source} & \textbf{Format String} & \textbf{Second-Order Files} & \textbf{Third-Order Files} \\
\hline
Numpy & \texttt{numpy} & \texttt{second.npy} & \texttt{third.npz} or \texttt{third.npy} \\
LAMMPS & \texttt{lammps} & \texttt{Dyn.form} & \texttt{THIRD} \\
{VASP + ShengBTE} & {\texttt{vasp-sheng}} & \texttt{FORCE\_CONSTANTS\_2ND} or \texttt{FORCE\_CONSTANTS} & \texttt{FORCE\_CONSTANTS\_3RD} \\
{QE + ShengBTE} \footnote{It supports loading force constants with mixed formats. Here, it means that the second-order force constant comes from Quantum Espresso and the third-order force comes from thirdorder.py in ShengBTE \cite{li2014shengbte}.} & {\texttt{qe-sheng}} & \texttt{espresso.ifc2} & \texttt{FORCE\_CONSTANTS\_3RD} \\
VASP + d3q\footnote{The second-order force constant comes from VASP and the third-order force constant comes from Thermal2 (D3Q)~\cite{paulatto2013anharmonic,fugallo_ab_2013}.} & \texttt{vasp-d3q} & \texttt{FORCE\_CONSTANTS\_2ND} or \texttt{FORCE\_CONSTANTS} & \texttt{FORCE\_CONSTANTS\_3RD\_D3Q} \\
QE + d3q & \texttt{qe-d3q} & \texttt{espresso.ifc2} & \texttt{FORCE\_CONSTANTS\_3RD\_D3Q} \\
HiPhive & \texttt{hiphive} & \texttt{model2.fcs} & \texttt{model3.fcs} \\
TDEP & \texttt{tdep} & \texttt{infile.forceconstant} & \texttt{infile.forceconstant\_thirdorder} \\
\hline
\end{tabular}
\end{minipage}
\end{table}

A key capability introduced in version 2.0 is a native support for temperature-dependent effective potentials (TDEP)~\cite{hellman_temperature-dependent_2013, knoop_tdeptemperature_2024, folkner2024elastic}. TDEP extracts renormalized force constants from molecular dynamics trajectories, capturing anharmonic effects and temperature-induced soft mode stabilization critical in materials undergoing structural phase transitions. The typical workflow involves: (i) running finite-temperature MD with an appropriate interatomic potential (DFT-based or MLP), (ii) processing the trajectory with TDEP tools to fit effective IFCs, and (iii) importing the resulting force constants into $\kappa$ALDo via the \texttt{tdep} format.

Integration with machine-learned potentials (MLPs) proceeds through two primary pathways: Direct ASE~\cite{larsen_atomic_2017} calculators wrapping MLP implementations or  LAMMPS PHONON package\cite{thompson2022lammps}, enabling force evaluations during finite-difference IFC computation. Alternatively, MLPs can be employed within GPUMD \cite{fan_gpumd_2022} and ASE for large-scale MD simulations, with forces extracted during post-processing via TDEP.
Supported MLPs include NEP~\cite{fan_neuroevolution_2021}, MACE~\cite{batatia2025foundation}, MatterSim~\cite{yang_mattersim_2024}, Orb~\cite{duschatko_orb_2024}, DeepMD~\cite{wang2018deepmd}, and others accessible through ASE or LAMMPS interfaces. 

For a system with $N_b$ atoms in the unit cell and $N$ atoms in the supercell representation, a second-order IFC tensor has a maximum size of $3N_b\times 3N$ elements and is always stored as a dense matrix. 
Third-order force constants are stored as sparse tensors (SciPy \texttt{COO} format) to manage memory for large systems, where the dense representation would require $3N_b\times(3N)^2$ elements to be stored. To achieve a sparse representation while retaining good accuracy, users can specify a \texttt{third\_energy\_threshold} to discard negligible matrix elements, 
Distance-based truncation via \texttt{distance\_threshold} further compresses the representation of the anharmonic IFC tensor by neglecting interactions beyond a specified cutoff radius. This approximation is valid for short-range interactions, and convergence needs to be verified for each system.

\subsection{Phonon Properties and Scattering}

The \texttt{Phonons} class computes vibrational properties and phonon scattering rates from force constants. Core functionality includes frequency and eigenvector calculation (Eq.~\ref{eq:eigenvalue_crystal}), group velocity evaluation via the Hellmann-Feynman theorem, and three-phonon scattering matrix construction.

\subsubsection{Group Velocities}

Phonon group velocity, governing energy transport, is computed as:
\begin{equation}
v_{\mu\alpha} = \frac{\partial \omega_\mu}{\partial q_\alpha} = \frac{1}{2\omega_\mu}\sum_{i\beta i'\beta'}\eta^*_{i\beta\mu}\frac{\partial D_{i\beta i'\beta'}(\mathbf{q})}{\partial q_\alpha}\eta_{i'\beta'\mu},
\label{eq:groupvel}
\end{equation}
implemented via finite-difference evaluation of $\partial D/\partial q_\alpha$ or analytic derivatives when available from IFC spatial Fourier transforms.

\subsubsection{Phonon Populations and Heat Capacity}

The equilibrium phonon population follows Bose-Einstein statistics:
\begin{equation}
n_\mu(T) = \frac{1}{e^{\hbar\omega_\mu/k_B T}-1},
\label{eq:bose}
\end{equation}
or in the classical limit ($k_B T \gg \hbar\omega_\mu$), $n_\mu^{\text{cl}} = k_B T/(\hbar\omega_\mu)$. The modal heat capacity is:
\begin{equation}
c_\mu = k_B\left(\frac{\hbar\omega_\mu}{k_B T}\right)^2\frac{e^{\hbar\omega_\mu/k_B T}}{(e^{\hbar\omega_\mu/k_B T}-1)^2} = \hbar\omega_\mu\frac{\partial n_\mu}{\partial T},
\label{eq:heatcap}
\end{equation}
reducing to $c_\mu^{\text{cl}} = k_B$ classically. The flag \texttt{is\_classic} controls which statistics are employed.

\subsubsection{Three-Phonon Scattering}

Anharmonic three-phonon scattering arises from the first (cubic) anharmonic term of the Taylor expansion of the potential/free energy. Fermi's golden rule yields scattering rates for processes $\mu \rightarrow \mu' + \mu''$ (phonon splitting, ``+'') and $\mu + \mu' \rightarrow \mu''$ (phonon merging, ``$-$''):
\begin{equation}
\Gamma^{\pm}_{\mu\mu'\mu''} = \frac{\hbar\pi}{8\omega_\mu\omega_{\mu'}\omega_{\mu''}}|\Phi^{\pm}_{\mu\mu'\mu''}|^2 \Delta^{\pm}_{\mu\mu'\mu''}(n_{\mu'}, n_{\mu''}),
\label{eq:gamma}
\end{equation}
where $\Phi^{\pm}_{\mu\mu'\mu''}$ is the projected third-order potential, and $\Delta^{\pm}$ encodes energy and momentum conservation, and thermal population factors:
\begin{align}
\Delta^{+}_{\mu\mu'\mu''} &= (n_{\mu'}-n_{\mu''})\delta(\omega_\mu - \omega_{\mu'} + \omega_{\mu''})\delta_{\mathbf{q}_\mu, \mathbf{q}_{\mu'} - \mathbf{q}_{\mu''} + \mathbf{G}}, \\
\Delta^{-}_{\mu\mu'\mu''} &= (n_{\mu'}+n_{\mu''}+1)\delta(\omega_\mu + \omega_{\mu'} - \omega_{\mu''})\delta_{\mathbf{q}_\mu + \mathbf{q}_{\mu'}, \mathbf{q}_{\mu''} + \mathbf{G}},
\end{align}
with reciprocal lattice vectors $\mathbf{G}$ allowing umklapp processes that conserve the crystal momentum as in $\mathbf{q}\pm \mathbf{q'} - \mathbf{q''} = \mathbf{G}$. 
The projected potential is:
\begin{equation}
\Phi_{\mu\mu'\mu''} = \sum_{i\alpha i'\alpha'i''\alpha''}\frac{\phi^{\prime\prime\prime}_{i\alpha i'\alpha'i''\alpha''}}{\sqrt{m_i m_{i'}m_{i''}}}\eta_{i\alpha\mu}\eta_{i'\alpha'\mu'}\eta_{i''\alpha''\mu''},
\label{eq:phi3_proj}
\end{equation}
requiring summation over all atoms and Cartesian directions, the most computationally demanding step in ALD calculations. Note that the projected potential is symmetric with respect to the process type: $\Phi^+_{\mu\mu'\mu''} = \Phi^-_{\mu\mu'\mu''} \equiv \Phi_{\mu\mu'\mu''}$, as the same matrix element governs both emission and absorption. The $\pm$ distinction in Eq.~\ref{eq:gamma} arises solely from the different energy-momentum conservation rules and population factors encoded in $\Delta^{\pm}$, not from different interaction strengths. Specifically, $|\Phi^{\pm}_{\mu\mu'\mu''}|^2$ represents the scattering transition probability, which is intrinsic to the three-phonon interaction and independent of whether the process is absorption or emission.

Energy conservation in the three-phonon scattering processes ($\omega_\mu \pm \omega_\mu' - \omega_\mu''=0$) is implemented by smearing the frequencies with finite-width distribution functions:
\begin{equation}
\delta(\omega) \approx \frac{1}{\pi}\frac{{\sigma/2}}{\omega^2 + {(\sigma/2)^2}}\quad\text{(Lorentzian)},\qquad
\delta(\omega) \approx \frac{1}{{\sqrt{\pi}\sigma}}e^{-\omega^2/{\sigma^2}}\quad\text{(Gaussian)},
\label{eq:broadening}
\end{equation}
with bandwidth $\sigma$ set {\it via} \texttt{third\_bandwidth} or calculated adaptively from phonon group velocity dispersions~\cite{li2014shengbte}. Both functions are normalized to unity over $(-\infty,+\infty)$; scattering contributions are evaluated only for mode pairs satisfying $|\Delta\omega| < 2\sigma$ Three broadening schemes are available via the \texttt{broadening\_shape} parameter: Lorentzian (\texttt{shape='lorentz'}), which has long tails and is exact for exponentially decaying correlations; Gaussian (\texttt{shape='gauss'}), the most common choice with smooth and faster-decaying tails; and triangular (\texttt{shape='triangle'}), which uses linear interpolation for fast computation at reduced accuracy, suitable for initial screening calculations.

The phonon lifetime (inverse bandwidth) is:
\begin{equation}
\tau_\mu = \frac{1}{2\Gamma_\mu} = \frac{1}{2\sum_{\mu'\mu''}(\Gamma^{+}_{\mu\mu'\mu''} + \Gamma^{-}_{\mu\mu'\mu''})}.
\label{eq:lifetime}
\end{equation}

\subsubsection{Computational Implementation}

The bottleneck in ALD simulations is calculating phonon coupling terms and scattering matrix elements, which require projecting third-order force constants onto phonon eigenstates (Eq.~\ref{eq:phi3_proj}). This projection scales as $\mathcal{O}((N_q \cdot 3N_b)^3)$, where $N_q$ is the number of $q$-points. $\kappa$ALDo~2.0 implements this algorithm as tensor multiplications: $N_q \times (3N_b)^2$ for BTE calculations (two $N_q \times 3N_b$ tensor multiplications).
Several optimization strategies are employed:
\begin{itemize}
\item \textbf{Sparse tensor operations:} Third-order IFCs are stored in coordinate (COO) format and processed using TensorFlow sparse kernels, avoiding dense instantiation.
\item \textbf{GPU acceleration:} Matrix multiplications in the projection are offloaded to GPU via TensorFlow when available.
\item \textbf{Selective calculation:} Scattering rates are computed only for physical modes (frequency above \texttt{min\_frequency}, below \texttt{max\_frequency}), excluding acoustic modes at $\Gamma$ and user-defined frequency ranges.
\item \textbf{Caching:} Intermediate results (\texttt{\_ps\_and\_gamma}, \texttt{\_sparse\_phase\_and\_potential}) are cached to disk in multiple formats (NumPy \texttt{.npz}, HDF5), avoiding recomputation in iterative workflows.
\end{itemize}

The user can control the phonon projection framework through \texttt{Phonons} class parameters: \texttt{storage} (memory management strategy), \texttt{broadening\_shape} (energy conservation kernel), \texttt{is\_balanced} ({symmetrizes population factors in three-phonon scattering rates to enforce detailed balance, improving convergence on coarse $q$-grids}), and \texttt{third\_bandwidth} (custom broadening width).

\subsection{Thermal Conductivity}

Given phonon frequencies, velocities, and scattering rates from the \texttt{Phonons} object, the \texttt{Conductivity} class solves for thermal transport using the BTE~\cite{peierls_kinetischen_1929,ziman2001electrons} or the QHGK formalism~\cite{isaeva_modeling_2019}.

\subsubsection{Boltzmann Transport Equation}

Under an applied temperature gradient $\nabla_\alpha T$, the phonon distribution deviates from equilibrium, with the deviation $\delta n_\mu$ determined by the linearized BTE:
\begin{equation}
\mathbf{v}_\mu \cdot \nabla T \frac{\partial n_\mu}{\partial T} = -\sum_{\mu'}\tilde{\Gamma}_{\mu\mu'}\delta n_{\mu'},
\label{eq:BTE}
\end{equation}
where the scattering matrix decomposes as $\tilde{\Gamma}_{\mu\mu'} = \delta_{\mu\mu'}\Gamma^0_\mu + \Gamma^1_{\mu\mu'}$, with the diagonal terms calculated as: 
\begin{equation}
\Gamma^0_\mu = \sum_{\mu'\mu''}(\Gamma^+_{\mu\mu'\mu''} + \Gamma^-_{\mu\mu'\mu''})  
\end{equation}
and the off-diagonal terms:
\begin{equation}
\Gamma^1_{\mu\mu'} = -\sum_{\mu''}\left[\Gamma^{+}_{\mu\mu'\mu''} + \Gamma^{-}_{\mu\mu'\mu''} + \Gamma^{+}_{\mu'\mu\mu''} + \Gamma^{-}_{\mu'\mu\mu''}\right].
\label{eq:gamma1}
\end{equation}
that captures mode-coupling. 
Normal processes ($\mathbf{G}=0$) redistribute energy without relaxing heat flux, while umklapp processes ($\mathbf{G}\neq 0$) provide thermal resistance. 

The BTE can be rewritten as a linear algebra system in which the phonon mean free paths $\lambda_{\mu'}$ are calculated as: 
\begin{equation}
    \mathbf{v}_\mu = \sum_{\mu'} \left( \delta_{\mu\mu'}\Gamma^0_\mu + \Gamma^1_{\mu\mu'} \right) \lambda_{\mu'}.
    \label{eq:invBTE}
\end{equation}
The thermal conductivity is then obtained as:
\begin{equation}
\kappa_{\alpha\beta} = \frac{1}{N_q V}\sum_{\mu}c_\mu \mathrm{v}_{\mu\alpha}\lambda_{\mu\beta}.
\label{eq:kappa_general}
\end{equation}
The phonon mean free paths can be expressed in terms of phonon lifetimes as $\lambda_{\mu\beta} =\mathrm{v}_{\mu\beta} \tau_{\mu\beta}$.

\noindent Three solution methods are implemented:\\
\textbf{(i) Relaxation Time Approximation (RTA).} Neglecting off-diagonal coupling ($\Gamma^1=0$), the mean free path simplifies to $\lambda_{\mu\alpha} = \tau_\mu v_{\mu\alpha}$ with $\tau_\mu = 1/\Gamma^0_\mu$, yielding:
\begin{equation}
\kappa^{\text{RTA}}_{\alpha\beta} = \frac{1}{N_q V}\sum_{\mu}c_\mu \mathrm{v}_{\mu\alpha}\tau_\mu \mathrm{v}_{\mu\beta} = \frac{1}{N_q V}\sum_{\mu}c_\mu \mathrm{v}_{\mu\alpha}\frac{1}{\Gamma^0_\mu} \mathrm{v}_{\mu\beta}.
\label{eq:rta}
\end{equation}
This approximation neglects momentum-conserving normal (N) processes. Whereas it is usually sufficiently accurate for systems with strong Umklapp scattering, it may underestimate the thermal conductivity by more than 50\% in materials where N-processes are significant~\cite{ward2009ab}.

\noindent \textbf{(ii) Self-Consistent Iterative Method (SC).} The full BTE (Eq.~\ref{eq:invBTE}) is solved iteratively~\cite{omini_beyond_1996,broido_lattice_2005}:
\begin{equation}
\lambda^{(n+1)}_{\mu\alpha} = \frac{v_{\mu\alpha}}{\Gamma^0_\mu} - \frac{1}{\Gamma^0_\mu}\sum_{\mu'}\Gamma^1_{\mu\mu'}\lambda^{(n)}_{\mu'\alpha},
\label{eq:sc_iter}
\end{equation}
initialized with RTA solution $\lambda^{(0)}_{\mu\alpha} = v_{\mu\alpha}/\Gamma^0_\mu$. Convergence is guaranteed when $\|\Gamma^1/\Gamma^0\|<1$~\cite{cepellotti_thermal_2016}. The user specifies the maximum number of iterations (\texttt{n\_iterations}) and convergence tolerance (\texttt{tolerance}) on the relative change in the isotropically-averaged thermal conductivity: $|\kappa^{(n+1)} - \kappa^{(n)}|/\kappa^{(n)} < \epsilon_{\text{tol}}$ between successive iterations, where $\kappa = \text{Tr}(\kappa_{\alpha\beta})/3$. This method balances accuracy and memory, avoiding explicit storage of the full scattering matrix.

\noindent \textbf{(iii) Full Matrix Inversion.} The exact solution of Eq.~\ref{eq:invBTE} can be obtained by inverting the scattering matrix.
\begin{equation}
\kappa_{\alpha\beta} = \frac{1}{N_q V}\sum_{\mu\mu'}c_\mu v_{\mu\alpha}(\tilde{\Gamma}^{-1})_{\mu\mu'}v_{\mu'\beta}.
\label{eq:inverse}
\end{equation}
This provides the most accurate BTE solution but requires storing the full $(N_q \cdot 3N) \times (N_q \cdot 3N)$ scattering matrix, limiting applicability to moderate system sizes.

Method selection is specified via \texttt{method='rta'/'sc'/'inverse'/'full'} when instantiating the \texttt{Conductivity} object. Memory scaling differs significantly: RTA and self-consistent methods require $N_q \times (3N)^2$ storage, while full inversion and eigendecomposition require $(N_q \cdot 3N)^2$ storage. For most bulk materials at moderate temperatures, the self-consistent method provides an optimal balance of accuracy and computational cost, while RTA is suitable for quick estimates and materials with strong anharmonicity.

\textbf{Finite-size effects, boundary scattering, and eigendecomposition approach.} 
When characteristic sample dimensions $L_\alpha$ become comparable to or smaller than phonon mean free paths, confinement and boundary scattering fundamentally alter thermal transport~\cite{shinde_length-scale_2014, benenti_non-fourier_2023}. This regime is ubiquitous in nanoscale systems, including nanowires, thin films, superlattices, and nanostructured thermoelectrics, as well as in low-dimensional materials (graphene, nanotubes) at cryogenic temperatures, where thermal phonon mean free paths can exceed millimeters~\cite{regner_broadband_2013}.

$\kappa$ALDo~2.0 provides two complementary approaches to address finite-size transport. The first adds a boundary scattering term to the diagonal of the full scattering matrix before inversion~\cite{maassen_simple_2015,kaiser_thermal_2017}:
\begin{equation}
\tilde{\Gamma}^{\text{fs}}_{\mu\mu'} = \tilde{\Gamma}_{\mu\mu'} + \delta_{\mu\mu'}\frac{2|v_{\mu\alpha}|}{L_\alpha},
\label{eq:finite_size}
\end{equation}
where $v_{\mu\alpha}$ is the group velocity component and $L_\alpha$ the sample dimension along direction $\alpha$. The mean free path is then obtained from full matrix inversion:
\begin{equation}
\lambda^{\text{fs}}_{\mu\alpha} = \sum_{\mu'}[(\tilde{\Gamma}^{\text{fs}})^{-1}]_{\mu\mu'}v_{\mu'\alpha}.
\label{eq:mfp_finite_size}
\end{equation}
Unlike Matthiessen's rule, which would simply add scattering rates for independent modes ($\tau^{-1}_{\text{eff}} = \tau^{-1}_{\text{anh}} + 2|v|/L$), this formulation preserves the off-diagonal scattering matrix elements $\Gamma^1_{\mu\mu'}$ that describe momentum exchange between modes, correctly capturing the interplay between boundary scattering and collective phonon transport.
This is selected via \texttt{finite\_length\_method='ms'}, with sample dimensions specified via \texttt{length=(L\_x, L\_y, L\_z)} in Ångströms (using \texttt{None} or 0 for infinite dimensions). For strongly confined systems with $L \ll \lambda_{\text{ph}}$, the ballistic limit (\texttt{finite\_length\_method='ballistic'}) is also available. 

The second approach employs full eigendecomposition of the scattering matrix (\texttt{method='full'})~\cite{cepellotti_thermal_2016}. The method diagonalizes the scattering matrix $\Gamma = U\Lambda U^T$, where $\Gamma$ is the $(N_q \cdot 3N) \times (N_q \cdot 3N)$ phonon-phonon scattering matrix, $U$ contains the eigenvectors (columns), and $\Lambda$ is a diagonal matrix of eigenvalues representing the decay rates of the collective heat-carrying excitations (termed "relaxons" in the original formulation). These collective modes decay exponentially as $e^{-\lambda_i t}$ with time constants $\tau_i = 1/\lambda_i$, where $\lambda_i$ are the eigenvalues. Rather than treating each phonon mode independently as in RTA, this approach identifies the true collective eigenmodes of the scattering operator that carry heat. This is essential in the hydrodynamic regime where momentum-conserving normal (N) processes dominate, causing breakdown of conventional RTA and self-consistent approximations. The eigendecomposition enables mode-by-mode analysis of transport contributions through the eigenvalue spectrum and can filter unphysical negative eigenvalues via \texttt{is\_using\_gamma\_tensor\_evects}, improving numerical stability. For a detailed description of the formalism and its physical interpretation, see Ref.~\cite{cepellotti_thermal_2016, cepellotti_boltzmann_2017}.

\subsubsection{Quasi-Harmonic Green-Kubo (QHGK)}
The QHGK approach~\cite{isaeva_modeling_2019} provides a unified lattice dynamic theory for both crystalline and disordered systems. This theory bridges and extends the BTE method for crystals and the Allen-Feldman theory for glasses~\cite{allen1989thermal,allen1993thermal}, and results in equations practically equivalent to the Wigner transport formalism~\cite{simoncelli_unified_2019,fiorentino_green-kubo_2023}, naturally capturing diffuson and locon contributions~\cite{allen1999diffusons,lundgren2021mode} (vibrational modes that transport heat through diffusion or remain localized, respectively).

The QHGK approach is particularly relevant for disordered, defective, and amorphous materials, where translational symmetry is broken, and the traditional BTE framework (which relies on well-defined wavevectors and Bloch states) becomes inapplicable, and for strongly anharmonic crystals for which the interband contribution to heat transport is substantial. 
For amorphous systems, QHGK is formulated as a real-space, $\Gamma-$point approach for large simulation cells ($N_q=1$, large $N$), using $(3N)^2$ storage to save the system state, making it computationally efficient for modeling disorder at the atomic scale. The thermal conductivity is:
\begin{equation}
\kappa_{\alpha\beta} = \frac{1}{V}\sum_{\mu\mu'}c_{\mu\mu'} v_{\mu\mu'}^{\alpha}\tau_{\mu\mu'} v_{\mu\mu'}^{\beta},
\label{eq:qhgk}
\end{equation}
where $c_{\mu\mu'}$ are generalized modal heat capacities defined as $c_{\mu\mu'} = \frac{k_B}{T}\frac{\omega_\mu\omega_{\mu'}(n_\mu - n_{\mu'})}{\omega_{\mu'}-\omega_\mu}$ for $\mu \neq \mu'$ and $c_{\mu\mu} = c_\mu$ (the diagonal heat capacity from Eq.~\ref{eq:heatcap}) for degenerate modes. The generalized velocities are defined as $v_{\mu\mu'}^{\alpha} = \sum_{i\beta i'\beta'}\eta^*_{i\beta\mu}\frac{\partial D_{i\beta i'\beta'}}{\partial R_{i'\alpha}}\eta_{i'\beta'\mu'}$, where $\partial D/\partial R_{i'\alpha}$ represents the derivative of the dynamical matrix with respect to atomic position, capturing how vibrational modes couple through atomic displacements. This operator captures non-diagonal contributions arising from the breakdown of wavevector definition in disordered systems, essential for representing diffuson transport. The generalized lifetimes $\tau_{\mu\mu'}$ are computed from thermal diffusivities with a broadening function. For Lorentzian broadening, 
\begin{equation}
    \tau_{\mu\mu'} = \frac{\gamma_{\mu\mu'}}{2[(\omega_\mu - \omega_{\mu'})^2 + \gamma_{\mu\mu'}^2]},
\end{equation}

where $\gamma_{\mu\mu'}$ is the bandwidth parameter controlling the energy resolution.

However, the impact of QHGK is not limited to disordered systems. Indeed, in crystals, the QHGK approach highlights an additional transport contribution on top of the BTE one. In the RTA, it reads:
\begin{equation}
\kappa_{\alpha\beta} =\kappa^{\mathrm{RTA}}+ \frac{1}{V}\sum_{\mathbf q s\neq s^\prime}c_{\mathbf q s s'} v_{\mathbf q s s^\prime}^{\alpha}\tau_{ \mathbf q s s'} v_{ \mathbf q s^\prime s}^{\beta},
\label{eq:qhgk_crystal}
\end{equation}
where phonon is explicitly labeled with its own Bloch wavevector $\mathbf{q}$ and band index $s$, $\mu=(\mathbf{q} s)$. The second sum in the expression indicated the so-called interband contribution~\cite{simoncelli_unified_2019}, between pairs of phonons with the same wavevector but different band indices. Although the BTE is generally dominant for most crystals, it has been shown that the additional term can significantly affect the thermal conductivity and its temperature dependence in certain crystals, e.g., GeTe and halide perovskites~\cite{dangic_origin_2021,simoncelli2022wigner}. Finally, the QHGK formalism can be further extended beyond the RTA by employing the full scattering matrix in the intraband term~\cite{simoncelli_unified_2019,fiorentino_green-kubo_2023}.
In practice, \(\kappa^{\mathrm{RTA}}\) is replaced by the full BTE result, while the interband contribution is retained at the RTA level.
As shown in Ref.~\cite{fiorentino_green-kubo_2023}, the full scattering matrix is typically relevant only at very low temperatures, where the interband contribution becomes negligible, thereby justifying the simplified RTA treatment adopted for the latter.

In addition to thermal conductivity, the QHGK method computes mode-resolved thermal diffusivities $D_\mu$ accessible via \texttt{Conductivity.diffusivity}, providing insight into the spatial extent of vibrational energy transport and enabling identification of propagons, diffusons, and locons~\cite{allen1999diffusons} in disordered materials. Implementation details follow Ref.~\cite{isaeva_modeling_2019}, with broadening controlled by \texttt{diffusivity\_bandwidth} and \texttt{diffusivity\_shape} (Lorentzian, Gaussian, or triangular kernels).

\subsection{\revtwo{Overview of New Capabilities in Version 2.0}}
\label{sec:overview}

\revtwo{Relative to the 2020 release~\cite{barbalinardo2020efficient}, $\kappa$ALDo~2.0 adds capabilities in three areas: the physics that can be modeled, the tools available to analyze it, and how efficiently the calculation runs.}

\revtwo{On the physical side, version 2.0 broadens the range of materials and properties accessible to lattice dynamics. The quasi-harmonic Green-Kubo solver treats disordered and strongly anharmonic systems on the same footing as crystals, capturing the interband transport that the Boltzmann equation alone omits. Isotopic mass disorder, non-analytical corrections for polar materials, and finite-size boundary scattering are available as modeling options. Elastic constants, the harmonic free energy, and thermal expansion through the quasi-harmonic approximation extend the code beyond transport into thermodynamic and mechanical properties.}

\revtwo{For analysis, version 2.0 adds diagnostics that characterize a material before and after a transport calculation. A scalar anharmonicity score measures how far a material departs from the harmonic approximation, which indicates whether perturbative methods are appropriate. The participation ratio measures how spatially localized each vibrational mode is, which helps to classify modes as propagons, diffusons, or locons~\cite{allen1999diffusons} in disordered systems.}

\revtwo{On the computational side, version 2.0 targets larger systems and shorter time to solution. The projection of third-order force constants onto phonon modes runs on the GPU through TensorFlow, and the temperature-dependent part of the calculation is separated from the temperature-independent part so that multi-temperature sweeps reuse the expensive projection. Finite-difference force-constant calculations and the per-$q$-point phonon calculation can be distributed across CPU cores through a shared-memory worker pool, with on-disk checkpointing so that interrupted runs resume rather than restart. Crystal symmetry is exploited to reduce the number of explicit IFC calculations during finite differences and to restrict the scattering calculation to the irreducible Brillouin zone, with the results mapped back onto the full grid, which lowers the cost of crystalline calculations without affecting the result. Modular storage backends keep the memory footprint manageable for systems of more than 10,000 atoms.}

\revtwo{The remainder of this section documents these capabilities in detail: Section~\ref{sec:modeling-extensions} covers the physical modeling options, Section~\ref{sec:analysis} the analysis and diagnostic tools, and Section~\ref{sec:architecture} the software architecture and performance features.}

\subsection{Modeling Extensions}
\label{sec:features}
\label{sec:modeling-extensions}

Beyond core thermal conductivity calculations, $\kappa$ALDo~2.0 provides specialized capabilities for challenging materials and computational efficiency enhancements for large-scale studies.

\subsubsection{Isotopic Scattering}

Natural isotopic disorder introduces mass variance that scatters phonons even in otherwise perfect harmonic crystals.
Following Tamura’s perturbation theory~\cite{tamura_isotope_1983}, the isotopic scattering rate for a phonon mode \((\mathbf{q},s)\) reads
\begin{equation}
\gamma^{\mathrm{iso}}_{\mathbf{q}s}=\frac{\pi}{2N_q}\,\omega_{\mathbf{q}s}^{2}\sum_{\mathbf{q}' s'}\delta\!\left(\omega_{\mathbf{q}s}-\omega_{\mathbf{q}' s'}\right)\sum_{i\alpha}g_i\left|\eta_{\mathbf{q}s}^{*,i\alpha}\eta_{\mathbf{q}' s'}^{i\alpha}\right|^{2},
\label{eq:isotope}
\end{equation}
where \(\mathbf{q}\) and \(s\) denote the phonon wavevector and band index, respectively, \(\alpha\) labels the Cartesian direction, and
\(g_i = \sum_{\mathrm{iso}} f_{i,\mathrm{iso}} (\Delta M_{i,\mathrm{iso}}/M_i)^2\)
is the relative mass-variance parameter of the \(i\)-th atom in the unit cell, defined in terms of the isotope fraction \(f_{i,\mathrm{iso}}\) and mass deviation \(\Delta M_{i,\mathrm{iso}}\).
In practice, isotopic scattering can be enabled by setting \texttt{include\_isotopes=True}.
If the input \texttt{g\_factor} array is omitted, natural isotopic abundances from the ASE database are automatically used to compute it.
Moreover, by tuning the \texttt{g\_factor}, Tamura’s formalism provides a perturbative description of mass-disorder scattering in alloyed systems, such as silicon--germanium alloys~\cite{garg2011role,fiorentino2025effects}.

Finally, the total scattering rate becomes $\gamma_\mu = \gamma^{\text{anh}}_\mu + \gamma^{\text{iso}}_\mu$.

\subsubsection{Non-Analytical Term Corrections}

In polar materials (ionic or covalent crystals with non-zero Born effective charges), long-range dipole-dipole interactions render the dynamical matrix non-analytic at $\Gamma$~\cite{gonze1997dynamical}. This causes directional dependence of the longitudinal optical (LO) and transverse optical (TO) mode splitting in the $\mathbf{q}\rightarrow 0$ limit. Neglecting this effect produces incorrect phonon dispersions near $\Gamma$ and erroneous thermal conductivities.

$\kappa$ALDo~2.0 implements the non-analytical correction (NAC) following Gonze and Lee:
\begin{equation}
D^{\text{NAC}}_{i\alpha i'\beta}(\mathbf{q}) = \frac{1}{\sqrt{m_i m_{i'}}}\frac{4\pi}{V_{\text{cell}}}\frac{(\sum_\gamma q_\gamma Z^*_{i,\gamma\alpha})(\sum_{\gamma'}q_{\gamma'}Z^*_{i',\gamma'\beta})}{\sum_{\alpha\beta}q_\alpha\epsilon^\infty_{\alpha\beta}q_\beta},
\label{eq:nac}
\end{equation}
where $Z^*_{i,\gamma\alpha}$ is the Born effective charge tensor and $\epsilon^\infty_{\alpha\beta}$ the high-frequency dielectric tensor. These quantities are obtained from DFPT calculations (Quantum ESPRESSO's \texttt{ph.x}) and imported via the \texttt{is\_nac=True} flag during \texttt{Phonons} initialization. The NAC term is added to $D(\mathbf{q})$ before diagonalization for all $\mathbf{q}$ near $\Gamma$. Figure~\ref{fig:mgo_dispersion} in Section~\ref{sec:applications} demonstrates the importance of NAC for polar oxides.

\subsubsection{Low-Dimensional Systems: Nanowires and Nanotubes}

$\kappa$ALDo~2.0 provides specialized support for quasi-one-dimensional systems such as nanowires and carbon nanotubes via the \texttt{is\_nw} flag. In these geometries, translational symmetry is broken in two directions while periodic boundary conditions apply along the transport axis. Setting \texttt{is\_nw=True} modifies the k-point sampling strategy to focus computational effort along the transport direction (e.g., \texttt{kpts=[1, 1, 11]} for z-axis transport), and adjusts the thermal conductivity normalization to account for the reduced cross-sectional area rather than three-dimensional volume.

This capability is particularly valuable for studying thermal transport in carbon nanotubes, semiconductor nanowires, and other quasi-1D structures \rev{where boundary scattering and reduced dimensionality dominate}. When combined with finite-size effects (\texttt{length} parameter), $\kappa$ALDo~2.0 enables prediction of length-dependent thermal conductivity in nanotubes and nanowires, critical for thermoelectric and thermal interface material applications~\cite{barbalinardo_ultrahigh_2021}.

\subsubsection{Elastic Properties}

The same second-order force constants used for phonon calculations also determine elastic properties. The elastic stiffness tensor $C_{ijkl}$ relates stress and strain via generalized Hooke's law. Following the derivation in Ref.~\cite{born1996dynamical,bornlongwave}, the elastic tensor is computed from:
\begin{equation}
C_{ijkl} = \frac{1}{V}\left[\sum_{nm}\phi^{(2)}_{nm,ik,jl} + \sum_{n\mu}\frac{(\partial D/\partial \epsilon_{ij})_{n\mu}(\partial D/\partial \epsilon_{kl})_{n\mu}}{\omega^2_\mu}\right],
\label{eq:elastic}
\end{equation}
where indices $i,j,k,l$ denote Cartesian directions and strain components, and $\epsilon$ is the strain tensor. The implementation in \texttt{ForceConstants.elastic\_prop()} returns the full fourth-rank tensor in GPa. Derived quantities, like the bulk modulus $B$, shear modulus $G$, and Young's modulus $E$, are extracted using Voigt-Reuss-Hill averaging. This capability is particularly valuable when combined with TDEP, enabling temperature-dependent elastic constant determination from the same MD trajectories used for thermal conductivity~\cite{folkner2024elastic}.

\subsubsection{Harmonic Free Energy and Thermal Expansion via Quasi-Harmonic Approximation}

$\kappa$ALDo~2.0 implements calculation of the harmonic contribution to the Helmholtz free energy for a system of harmonic oscillators with frequencies $\omega_\mu(\mathbf{q})$:
\begin{equation}
F_{\text{vib}}(T) = \sum_{\mathbf{q}\mu}\left[\frac{\hbar\omega_\mu(\mathbf{q})}{2} + k_B T\ln\left(1-e^{-\hbar\omega_\mu(\mathbf{q})/(k_B T)}\right)\right],
\label{eq:harmonic_free_energy}
\end{equation}
which includes both the zero-point energy contribution and the thermal population term. This quantity is essential for thermodynamic calculations and enables thermal expansion predictions via the quasi-harmonic approximation (QHA)~\cite{biernacki_negative_1989, fleszar_first-principles_1990, baroni_thermal_2010}.

In the QHA framework, temperature-dependent equilibrium volumes and elastic properties are determined by minimizing the total Helmholtz free energy:
\begin{equation}
F(V,T) = E_0(V) + F_{\text{vib}}(V,T) = E_0(V) + \sum_\mu\left[\frac{\hbar\omega_\mu(V)}{2} + k_B T\ln\left(1-e^{-\hbar\omega_\mu(V)/(k_B T)}\right)\right],
\label{eq:qha}
\end{equation}
where $E_0(V)$ is the static lattice energy and $F_{\text{vib}}(V,T)$ is the vibrational free energy evaluated at different volumes. By calculating force constants and phonon frequencies across a grid of lattice parameters (controlled via the \texttt{quasiharmonic} module), the code determines equilibrium volumes, thermal expansion coefficients $\alpha(T) = (1/V)(\partial V/\partial T)_P$, and pressure-temperature phase diagrams. This functionality complements TDEP workflows by providing an alternative route to finite-temperature lattice properties when the computational cost of systematic volume sampling is acceptable. QHA is particularly useful for materials with moderate anharmonicity where explicit inclusion of temperature-dependent force renormalization (via TDEP) may not be necessary.

\subsection{Analysis Tools}
\label{sec:analysis}

\subsubsection{Anharmonicity Quantification}

Before devoting to expensive thermal transport calculations, it is useful to assess whether anharmonicity is significant. The $\sigma_A$ score~\cite{knoop2020anharmonicity} represents anharmonic strength by comparing actual forces to their anharmonic contributions:
\begin{equation}
\sigma_A = \sqrt{\frac{\langle(F^{\text{anharm}})^2\rangle}{\langle F^2\rangle}},  \ F^{\text{anharm}} = F - F^{\text{harm}},
\label{eq:sigma_a}
\end{equation}
where $F$ are forces along an MD trajectory and $F^{\text{harm}} = -\Phi^{(2)} \cdot u$ are forces predicted from second-order IFCs and displacements $u$. $\sigma_A $ below 0.1 indicates a harmonic regime, $0.1 < \sigma_A < 0.5$ for weak anharmonicity, and $\sigma_A$ greater than 0.5 signifies strong anharmonicity, which indicates that accurate lattice thermal conductivity predictions might require beyond three-phonon BTE or molecular dynamics. The function \texttt{sigma2\_tdep\_md()} in the \texttt{controllers.sigma2} module automates this calculation given TDEP force constants and an MD trajectory.

\subsubsection{Participation Ratio}

The participation ratio $PR_\mu$ quantifies mode localization, particularly important in disordered systems:
\begin{equation}
PR_\mu = \frac{1}{N}\left(\sum_{i=1}^{N}\sum_{\alpha}|\eta_{i\alpha\mu}|^2\right)^2\left(\sum_{i=1}^{N}\sum_{\alpha}|\eta_{i\alpha\mu}|^4\right)^{-1}.
\label{eq:participation}
\end{equation}
Extended modes have $PR \sim 1$, while localized modes have $PR \ll 1$~\cite{allen1999diffusons}. Computed automatically via \texttt{Phonons.participation\_ratio}, this property helps identify diffuson and locon contributions to thermal conductivity.

\subsection{Software Architecture and Performance}
\label{sec:architecture}

\subsubsection{Object-Oriented Design and Lazy Evaluation}

$\kappa$ALDo 2.0 employs a three-tier class hierarchy mirroring the computational workflow: \texttt{ForceConstants} manages IFCs and atomic geometry, \texttt{Phonons} computes vibrational properties and scattering rates, and \texttt{Conductivity} solves the BTE or QHGK. Each class inherits from the \texttt{Storable} base class, implementing lazy evaluation via the \texttt{@lazy\_property} decorator pattern. Expensive properties (e.g., scattering rates, mean free paths) are computed only when accessed and cached automatically based on a label encoding relevant parameters (temperature, $q$-mesh, statistics). This dramatically reduces computation time for iterative workflows like convergence testing and parameter sweeps.

\subsubsection{Storage Backends}

Four storage formats accommodate different use cases, selected via \\
\texttt{storage='formatted'/'numpy'/'hdf5'/'memory'}: formatted text files enable human inspection and interoperability; binary NumPy arrays provide fast I/O for production calculations; HDF5 organizes complex multi-parameter datasets; and memory-only mode suits transient calculations. Table~\ref{tab:storage} provides detailed performance characteristics.

\begin{table}[h!]
\centering
\caption{Comparison of storage backends for caching intermediate results in $\kappa$ALDo~2.0 calculations.}
\label{tab:storage}
\begin{tabular}{p{2.2cm}p{3.5cm}p{3cm}p{4.5cm}}
\hline
\textbf{Backend} & \textbf{I/O Speed} & \textbf{Storage Efficiency} & \textbf{Recommended Use Case} \\
\hline
\texttt{formatted} & Slow (text parsing) & Low (ASCII) & Human inspection; small systems; sharing with non-Python tools \\
\texttt{numpy} & Fast (binary) & High (compressed) & Production calculations; large systems; iterative workflows \\
\texttt{hdf5} & Fast (binary) & High (compressed) & Complex datasets; multiple parameter sets; database integration \\
\texttt{memory} & Fastest (RAM only) & N/A (no disk) & Transient calculations; HPC job arrays; testing \\
\hline
\end{tabular}
\end{table}

\subsubsection{Visualization and Analysis Tools}

$\kappa$ALDo~2.0 includes comprehensive plotting utilities via the \texttt{plotter} module for rapid visualization and analysis of phonon properties. The \texttt{plot\_crystal()} function automatically generates publication-quality figures for crystalline materials including: (i) phonon dispersion along high-symmetry paths, (ii) phonon density of states, (iii) group velocity versus $q$-vector, (iv) group velocity versus frequency, (v) heat capacity versus frequency, (vi) three-phonon scattering phase space, (vii) phonon lifetimes, (viii) scattering rates, (ix) mean free paths, (x) per-mode thermal conductivity contributions, (xi) cumulative $\kappa$ versus frequency, and (xii) cumulative $\kappa$ versus mean free path. For disordered and amorphous systems, \texttt{plot\_amorphous()} provides all frequency-dependent plots plus mode thermal diffusivity and participation ratio analyses for identifying propagons, diffusons, and locons. Individual plotting functions (\texttt{plot\_dispersion()}, \texttt{plot\_dos()}, \texttt{plot\_vs\_frequency()}) enable custom analysis workflows. All figures are automatically saved in high resolution with systematic folder organization based on simulation parameters (temperature, $q$-mesh, solution method), facilitating comparison across parameter sets.

\subsubsection{Performance Optimization}

Key performance features include:

\textbf{Sparse tensor operations:} Third-order IFCs and scattering matrices are stored and manipulated in sparse formats (COO, CSR), reducing memory footprint by 100–1000$\times$ for typical systems.

\begin{figure}[h!]
    \centering
    \includegraphics[width=0.7\linewidth]{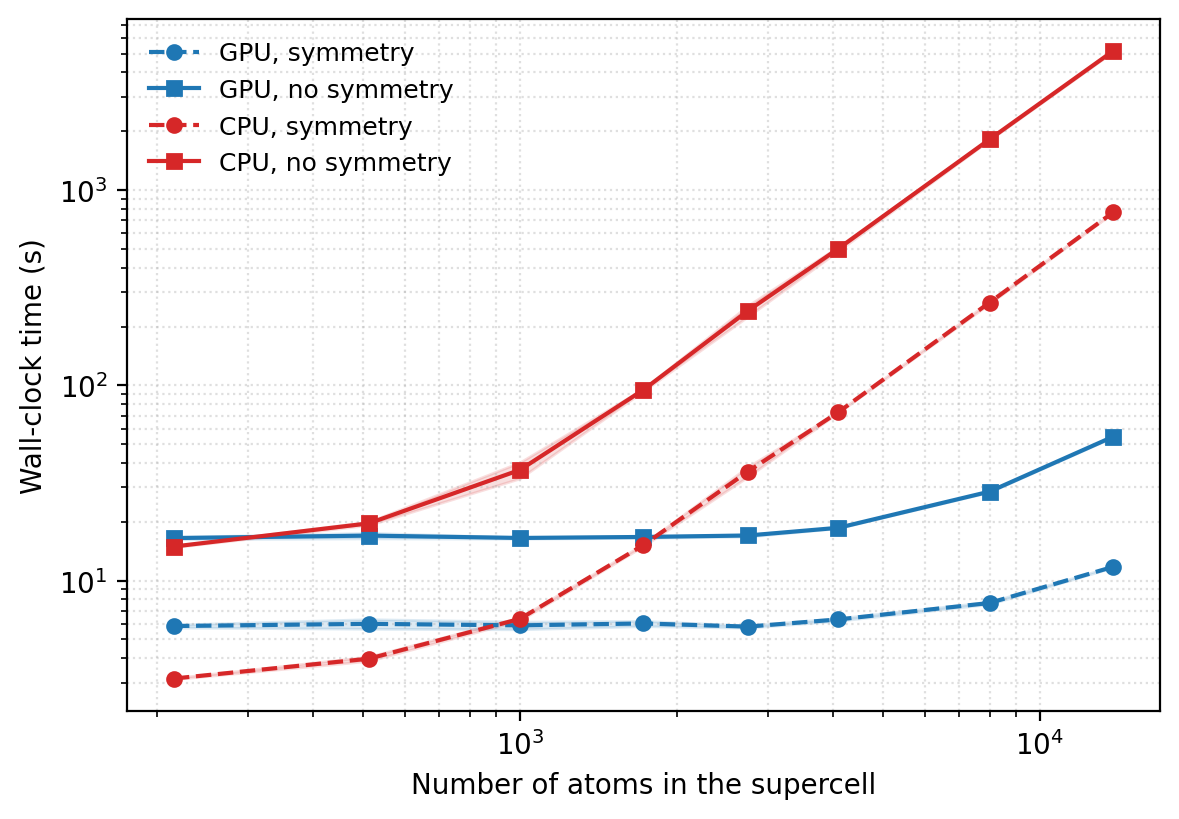}
    \caption{\revtwo{Runtime of QHGK lattice thermal conductivity calculations for bulk \ce{Si} at \SI{300}{\kelvin} as a function of supercell size, at a fixed $3\times3\times3$ $q$-grid, with interatomic force constants from the Tersoff potential~\cite{tersoff1988new}. Each curve compares the GPU and CPU code paths, with and without crystal-symmetry reduction of the Brillouin-zone sampling. Benchmarks were run on a single NVIDIA H200 GPU and on an Intel Xeon Platinum 8580 node (about eleven CPU cores were used by the CPU runs). Every point is the mean of three runs; shaded bands show the standard deviation. The GPU path with symmetry stays close to flat across the range, reaching roughly 12~s at 13{,}824 atoms, while the CPU path without symmetry grows to several thousand seconds over the same range. This comparison reflects the relative scaling of identical double-precision code paths on two different architectures.}}
    \label{fig:run_time_analysis}
\end{figure}

\textbf{GPU acceleration:} TensorFlow backend automatically utilizes available GPUs for tensor contractions in the IFC projection step (Eq.~\ref{eq:phi3_proj}). \revtwo{The contractions are carried out in double precision (\texttt{complex128}/\texttt{float64}) on both CPU and GPU, so enabling the GPU changes the runtime, not the computed values. Figure~\ref{fig:run_time_analysis} reports the runtime as a function of system size, and shows that the GPU path keeps the calculation nearly flat in cost over the full range of sizes tested, whereas the CPU cost grows steeply.}

\textbf{Multi-temperature optimization:} Version 2.0 decouples temperature-invariant computations from temperature-dependent population factors, enabling efficient thermal conductivity calculations across multiple temperatures. The computationally intensive projection of third-order force constants onto phonon modes (Eq.~\ref{eq:phi3_proj}) is decoupled from temperature-dependent population factors. This allows the projection tensors to be computed once and cached, reducing the complexity of subsequent temperature sweeps from $\mathcal{O}(N_T \cdot N^3)$ to $\mathcal{O}(N^3 + N_T \cdot N^2)$, where $N_T$ is the number of temperatures, providing order-of-magnitude speedups for typical material screening workflows.

\textbf{Vectorization:} NumPy broadcasting and \texttt{opt\_einsum} optimized contractions minimize Python-level loops in scattering calculations.

\textbf{Parallel $q$-point and force-constant loops:} \revtwo{Phonon calculations at different $q$-points, and the finite-difference evaluations used to build the second- and third-order force constants, are independent across work items. Version 2.0 distributes them across CPU cores through a shared-memory worker pool (selected with \texttt{n\_workers}), built on Python \texttt{ProcessPoolExecutor}. Completed work items are checkpointed to disk, so an interrupted calculation resumes from where it stopped instead of restarting. The pool uses the \texttt{spawn} start method, which keeps it compatible with GPU-backed and TensorFlow-backed force calculators.}

\subsubsection{Deployment and Documentation}

$\kappa$ALDo 2.0 emphasizes reproducibility and accessibility:

\textbf{Containerization:} Docker images with all dependencies pre-installed enable one-line deployment across platforms.

\textbf{Cloud tutorials:} Google Colab notebooks provide interactive tutorials runnable in web browsers without local installation.

\textbf{Continuous integration:} Automated testing via CircleCI ensures code quality. Pull requests require passing 40+ unit tests covering core functionality and regression prevention.

\textbf{Auto-generated documentation:} Sphinx processes docstrings into HTML/PDF manuals hosted at \url{https://nanotheorygroup.github.io/kaldo}, synchronized with each release.

\revtwo{Taken together, these pieces address a recurring difficulty with research codes, namely that a calculation which runs on the author's machine is often hard to reproduce elsewhere. Pinning the full dependency stack in a container image removes version drift between machines, which matters in particular for the GPU stack, where the TensorFlow, CUDA, and driver versions have to agree. Running the test suite automatically on every pull request catches regressions before they reach users, and the same suite doubles as a set of worked reference calculations. Building the documentation from the docstrings on each release keeps the manual consistent with the code that ships. The main lesson from assembling this infrastructure was that the GPU-enabled image is the fragile part: it has to be rebuilt and re-tested whenever the upstream CUDA or TensorFlow versions move, and treating that image as a tested artifact rather than a convenience has been worth the effort.}

\textbf{Example library:} A standalone repository hosted at \url{https://github.com/nanotheorygroup/kaldo-examples} containing detailed examples to model lattice thermal conductivity of semiconducting materials (e.g.\ \ce{Si}, \ce{Ge}, \ce{SiC}, \ce{GaAs}, \ce{MgO}, \ce{AlN}, \ce{NaCl}) using various workflows including DFT with DFPT/D3Q, empirical potentials via LAMMPS, and machine-learned potentials (MatterSim~\cite{yang_mattersim_2024}, ORB~\cite{duschatko_orb_2024}, NEP~\cite{fan_neuroevolution_2021}, ACE~\cite{folkner2024elastic}, UPET~\cite{bigi2026pushing}).

\subsubsection{Example: Basic Workflow}

To illustrate typical usage, we present an example calculating the thermal conductivity of SiC using the MatterSim machine-learned potential with structure optimization:

\begin{lstlisting}
# Import kALDo classes and ASE
from kaldo.forceconstants import ForceConstants
from kaldo.phonons import Phonons
from kaldo.conductivity import Conductivity
import kaldo.controllers.plotter as plotter
from ase.build import bulk
from ase.optimize import BFGS
from ase.constraints import StrainFilter
from mattersim.forcefield import MatterSimCalculator

# Stage 1: Structure optimization
atoms = bulk('SiC', 'zincblende', a=4.35)
calc = MatterSimCalculator(device='cuda')
atoms.calc = calc

# Optimize lattice parameters and atomic positions
sf = StrainFilter(atoms)
opt = BFGS(sf)
opt.run(fmax=0.001)

# Stage 2: Compute force constants using finite differences
fc = ForceConstants(
    atoms=atoms,
    supercell=[10, 10, 10],
    third_supercell=[5, 5, 5],
    folder='fd_SiC_MatterSim'
)
fc.second.calculate(calc, delta_shift=0.03)
fc.third.calculate(calc, delta_shift=0.03)

# Stage 3: Calculate phonon properties
phonons = Phonons(
    forceconstants=fc,
    kpts=[15, 15, 15],
    temperature=300,
    is_classic=False,
    folder='ALD_SiC_MatterSim'
)

# Plot phonon dispersion
plotter.plot_dispersion(phonons, n_k_points=300)

# Stage 4: Calculate thermal conductivity
cond = Conductivity(phonons=phonons, method='inverse')

# Access results
kappa = cond.conductivity.sum(axis=0)
print(f"Thermal conductivity: {kappa.trace()/3:.1f} W/m/K")
\end{lstlisting}

More complex workflows involving TDEP, isotopic scattering, or NAC corrections follow this same pattern with additional keyword arguments documented in the API reference.

\subsubsection{Command-Line Interface}

$\kappa$ALDo~2.0 includes a command-line interface (\texttt{kaldo/cli/}) that accepts JSON-formatted configuration files, simplifying deployment on clusters and HPC systems. This declarative approach is particularly useful for batch job submission, parameter sweeps, and reproducible workflows where separating calculation parameters from execution logic is beneficial. A typical configuration file specifies the force constants, phonon calculation parameters, conductivity method, and optional visualization settings. While the Python API offers maximum flexibility for power users and complex research workflows, the JSON interface is optimized for high-throughput batch processing on HPC resources.

\lstset{language=json, numbers=none}
\begin{lstlisting}
{
  "description": "Example kALDo configuration using silicon crystal test data",
  "forceconstants": {
    "folder": "kaldo/tests/si-crystal",
    "format": "eskm",
    "supercell": [3, 3, 3]
  },
  "phonons": {
    "temperature": 300.0,
    "kpts": [5, 5, 5],
    "is_classic": false,
    "storage": "memory"
  },
  "conductivity": {
    "method": "rta",
    "storage": "memory"
  },
  "plotter": {
    "plot_dispersion": {
      "n_k_points": 300,
      "is_showing": false,
      "with_velocity": true,
      "folder": "plots"
    }
  }
}
\end{lstlisting}
\lstset{language=Python, numbers=left}

The configuration is executed via \texttt{kaldo < example.in}, making it ideal for automated workflows and ensuring complete reproducibility of calculations.

\section{Applications}
\label{sec:applications}

We demonstrate $\kappa$ALDo~2.0's advanced capabilities through case studies of two prototypical materials: the halide perovskite \ce{CsPbBr3} and the polar oxide \ce{MgO}. These examples showcase the software's key features, including TDEP-based temperature-dependent force constants for materials with strong anharmonicity (\ce{CsPbBr3}), non-analytical corrections for polar materials (\ce{MgO}), and seamless integration with machine-learned potentials (MatterSim) and electronic structure package (d3q). Together, these applications illustrate how $\kappa$ALDo~2.0 addresses challenging materials where conventional harmonic approaches fail, while providing the flexibility to combine multiple computational methods (DFPT, MLPs, TDEP) within a unified workflow.

\subsection{Cubic \texorpdfstring{\ce{CsPbBr3}}{CsPbBr3} using TDEP}

To study the cubic phase of \ce{CsPbBr3}, we employ the temperature-dependent effective potential (TDEP) methodology. Experimentally, \ce{CsPbBr3} transforms into the $Pm\bar{3}m$ phase above approximately \SI{411}{\kelvin}, where anharmonic lattice dynamics play a central role in stabilizing the structure. Within the harmonic approximation, the cubic phase is dynamically unstable due to soft phonon modes at the $M$ and $R$ points. By fitting temperature-dependent interatomic force constants from classical molecular dynamics trajectories, TDEP provides renormalized phonon dispersions that recover phase stability and enable a quantitative description of vibrational and thermal transport properties. In this phase, \ce{CsPbBr3} exhibits the highest lattice thermal conductivity among its polymorphs, although phonon spectral functions show strongly overdamped, non-Lorentzian lineshapes that signal intense anharmonicity.

To perform this calculation, we run an NPT equilibration step on a 6x6x6 cubic \ce{CsPbBr3} supercell followed by an NVT simulation in GPUMD using a NEP potential developed by Erik Fransson et al\cite{franssonCPBnep}. We sample 200 frames evenly from the trajectory and fit the TDEP second- and third-order interatomic force constants to the chosen frames. Thermal conductivity is then calculated on a 12x12x12 $q$-grid. {Non-analytical corrections are not applied in this example, as the LO--TO splitting is negligible in \ce{CsPbBr3} and has no effect on $\kappa$}~\cite{dangic_lattice_2025}.

\begin{figure}[h!]
    \centering
    \includegraphics[width=0.9\linewidth]{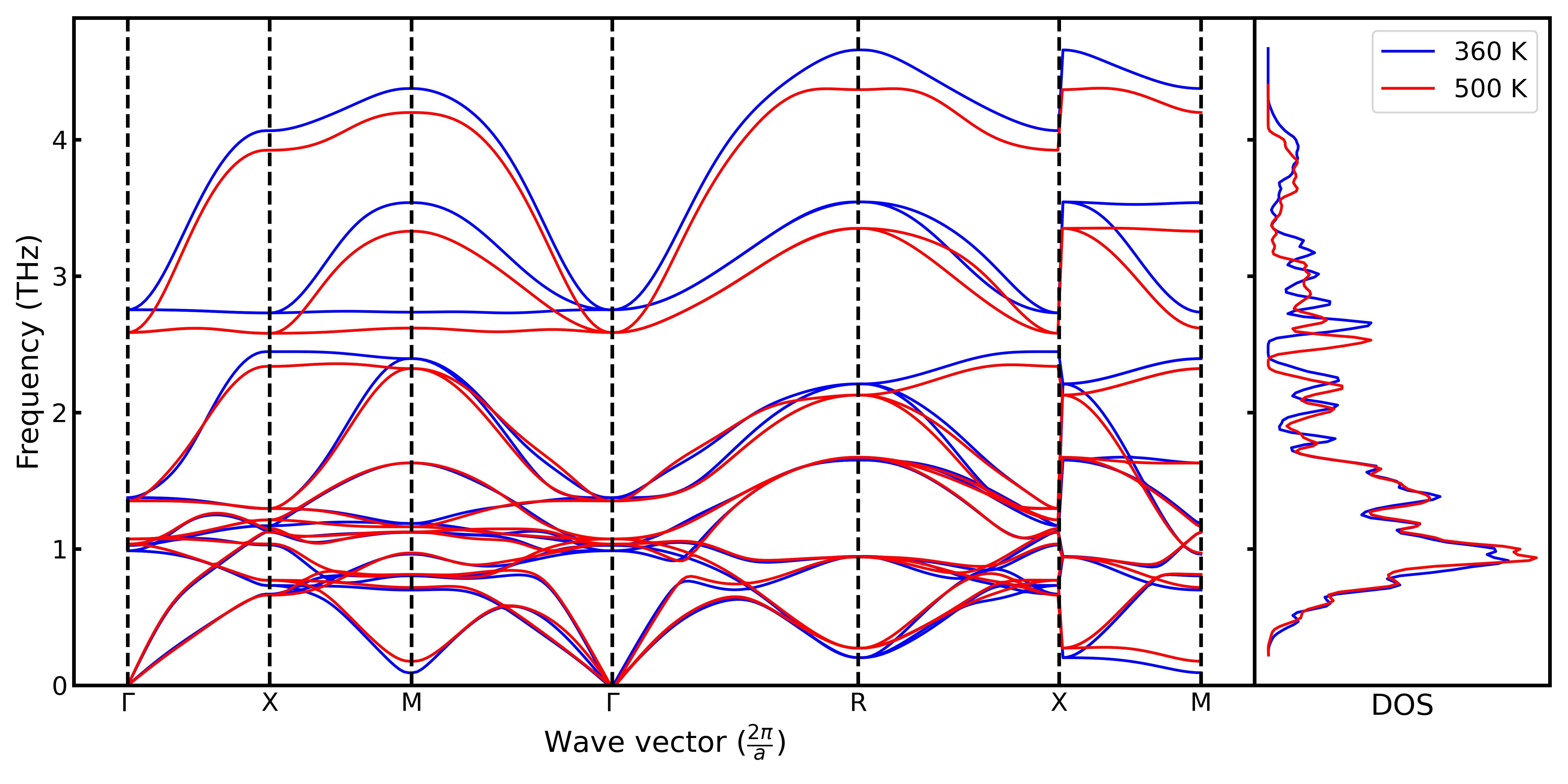}
    \caption{Phonon dispersion relation (left) and density of states (right) for cubic phase \ce{CsPbBr3} at 360 and \SI{500}{\kelvin}.}
    \label{fig:cpb_dispersion}
\end{figure}

\begin{figure}[h!]
    \centering
    \includegraphics[width=0.65\linewidth]{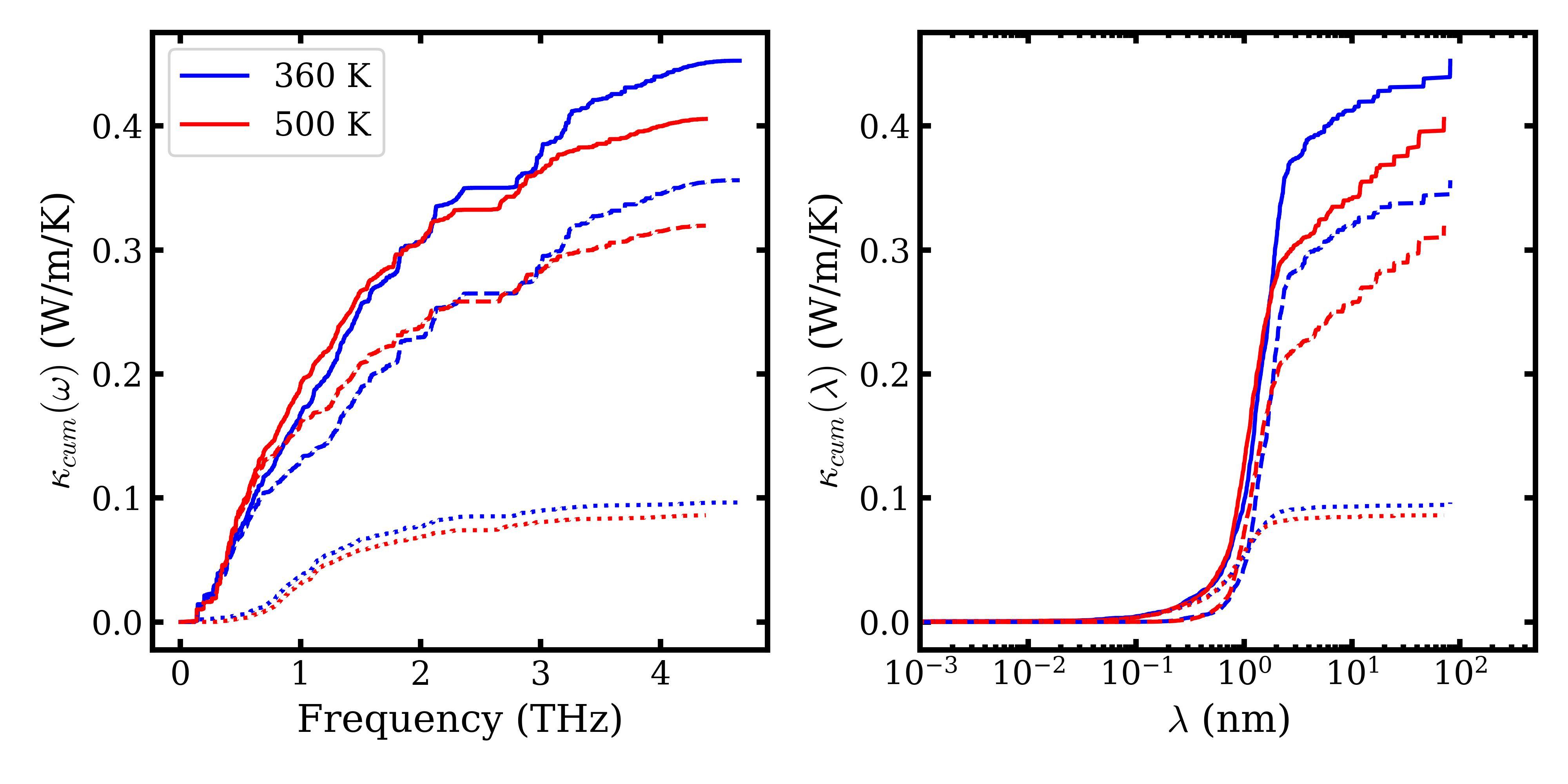}
    \caption{Cumulative thermal conductivity with respect to phonon mode frequency (left) and phonon mean free path (right). Dashed lines are used to indicate coherent contribution from QHGK, dotted lines are used to indicate the BTE contribution, and solid lines indicate total thermal conductivity calculated by adding the coherent (diffuson) contribution, estimated as the difference between the full QHGK and RTA, to the exact BTE solution ($\kappa_{inv}$).}
    \label{fig:cpb_kappa_cum_freq_mfp}
\end{figure}

In Fig. \ref{fig:cpb_dispersion}, the phonon dispersion relations capture the phonon mode softening at the M and R points as the phase transition of the NEP potential is approached from above ($\approx \SI{340}{\kelvin}$). The total impact of coherent transport is determined by calculating the interband phonon contribution to the thermal conductivity using {Eq.~\ref{eq:qhgk_crystal}, which gives the relation}~\cite{fiorentino_green-kubo_2023}:
\begin{equation}
    \kappa_{T} = \kappa_{inv} + (\kappa_{QHGK} - \kappa_{RTA})
    \label{eq:coherent_kappa}
\end{equation}
Interband transport contributions are present and account for $\approx 25\%$ of total thermal conductivity in Fig \ref{fig:cpb_kappa_cum_freq_mfp}, but they are relatively weaker than in the lower-symmetry tetragonal and orthorhombic phases, in line with experimental evidence of anomalous heat conduction in this material.

\begin{figure}[h!]
    \centering
    \includegraphics[width=0.9\linewidth]{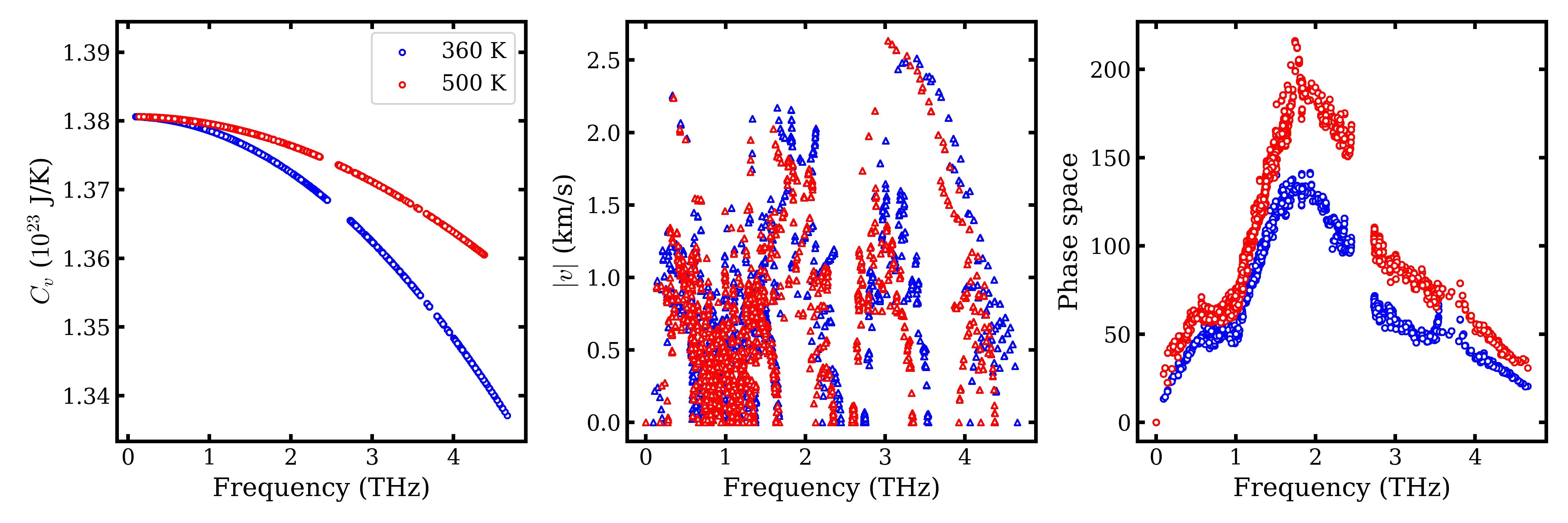}
    \caption{Phonon heat capacity (left), group velocity magnitude (middle), and scattering phase space (right) for 360 and \SI{500}{\kelvin} cubic \ce{CsPbBr3}.}
    \label{fig:cpb_cv_velocity_ps}
\end{figure}

\begin{figure}[h!]
    \centering
    \includegraphics[width=0.9\linewidth]{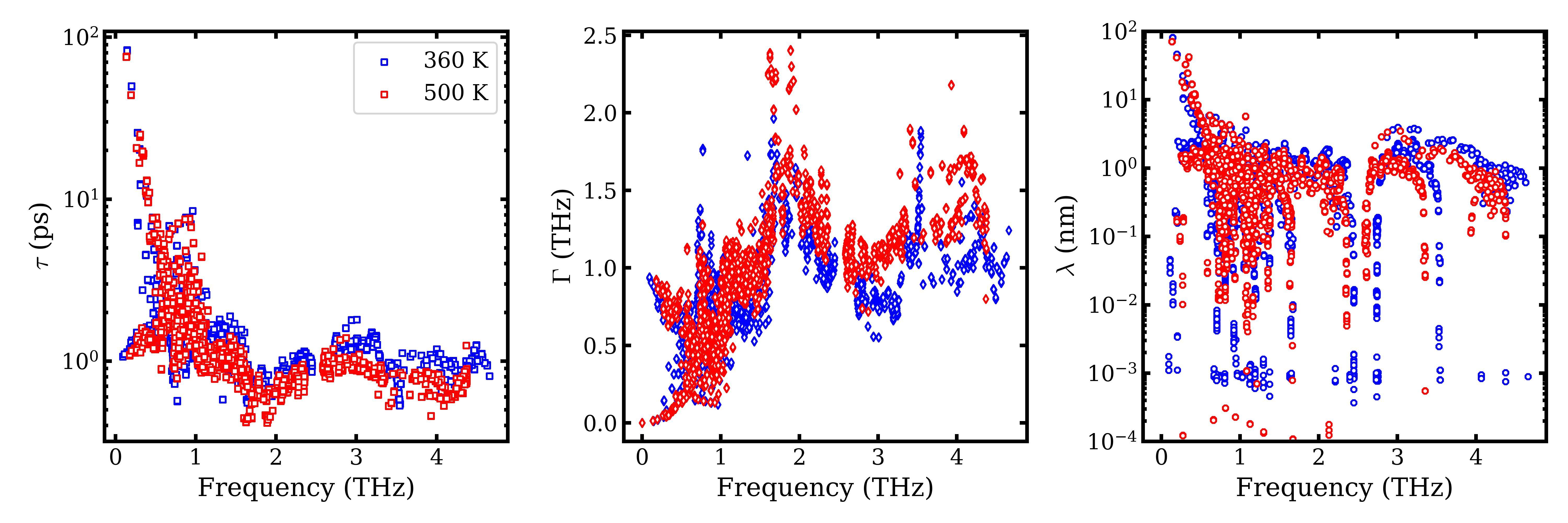}
    \caption{Phonon lifetimes (left), scattering rate (middle), and mean free path (right) for 360 and \SI{500}{\kelvin} cubic \ce{CsPbBr3}.}
    \label{fig:cpb_lifetime_rates_mfp}
\end{figure}

Lastly, we plot various properties related to the second and third order interatomic force constants in Fig. \ref{fig:cpb_cv_velocity_ps} and Fig. \ref{fig:cpb_lifetime_rates_mfp} to gauge the impact of finite temperature renormalization on them. We notice little impact on increasing temperature on second order properties from \SI{360}{\kelvin} to \SI{500}{\kelvin}, however, third order properties change much more dramatically, with a scattering phase space increase of $\approx 60\%$ and consistently larger lifetimes at \SI{360}{\kelvin} compared to \SI{500}{\kelvin} by about 1 order of magnitude.

\subsection{Thermal Properties of \texorpdfstring{\ce{MgO}}{MgO}: QHA Thermal Expansion and NAC Corrections}

To test our new free energy calculator and quasiharmonic algorithm, we perform the quasiharmonic approximation on \ce{MgO} from 0 to \SI{900}{\kelvin} using the MatterSim potential developed using PBE level of theory on a 3$\times$3$\times$3 supercell and 12$\times$12$\times$12 $q$-grid using 21 evenly spaced lattice constants between 4.23027 and 4.35846 \r{A}. The resulting free energy, lattice constant, and volumetric thermal expansion are shown below in Fig. \ref{fig:qha_mgo}.

\begin{figure}[h!]
    \centering
    \includegraphics[width=0.9\textwidth]{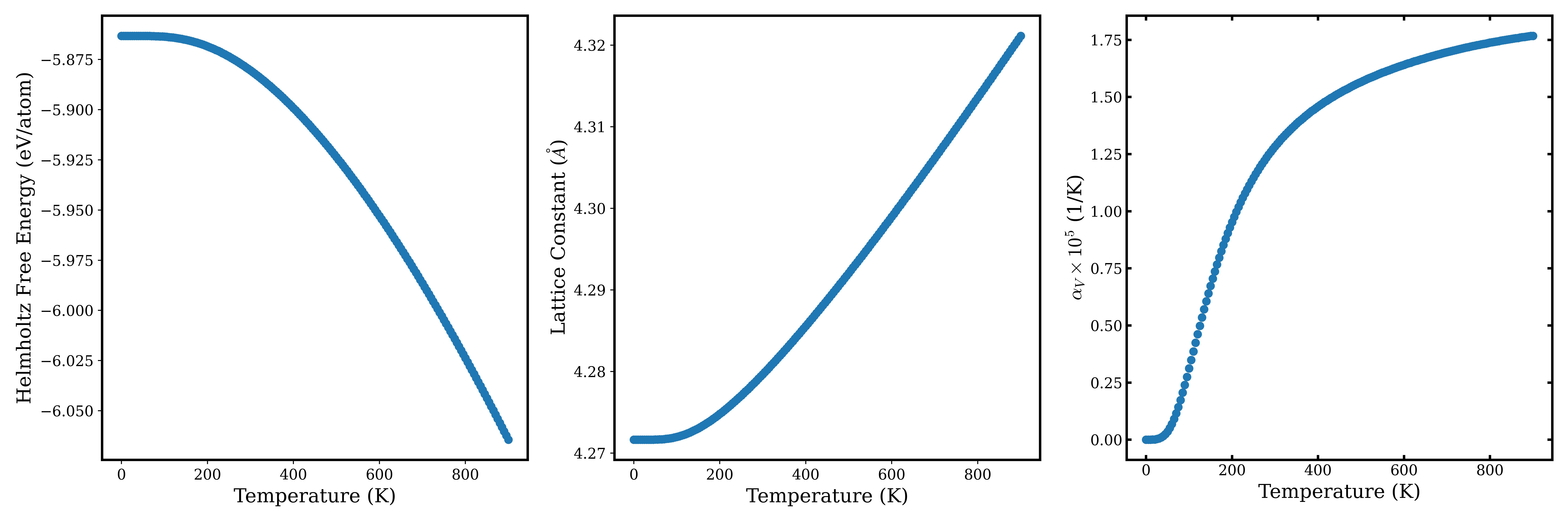}
    \caption{Quasiharmonic calculation performed on cubic phase \ce{MgO} using MatterSim machine learned potential.}
    \label{fig:qha_mgo}
\end{figure}

\begin{figure}[h!]
    \centering
    \includegraphics[width=0.75\textwidth]{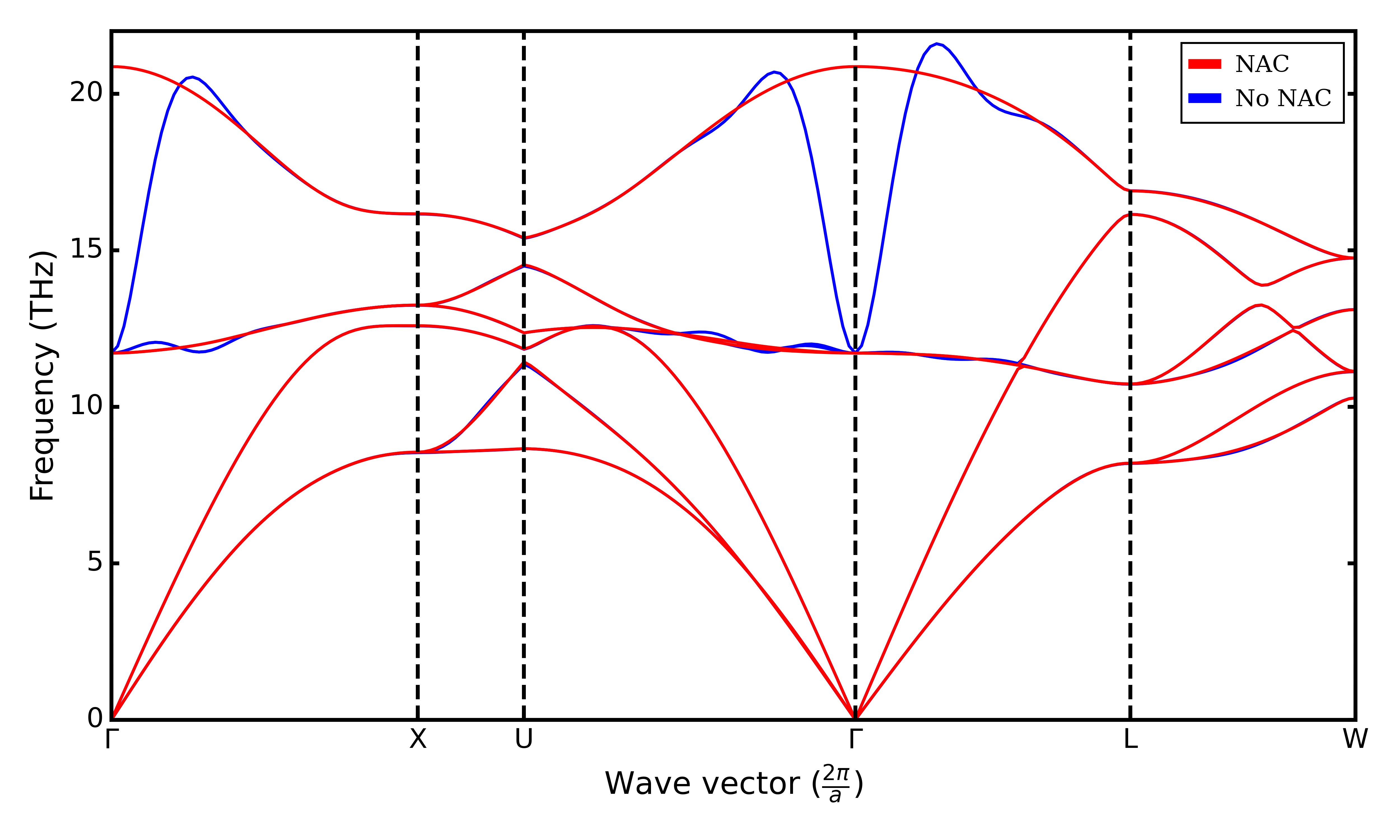}
    \caption{The phonon dispersion of  \ce{MgO} along a high symmetry path with (red) and without (blue) the NAC correction.}
    \label{fig:mgo_dispersion}
\end{figure}

\begin{figure}[h!]
    \centering
    \includegraphics[width=0.9\textwidth]{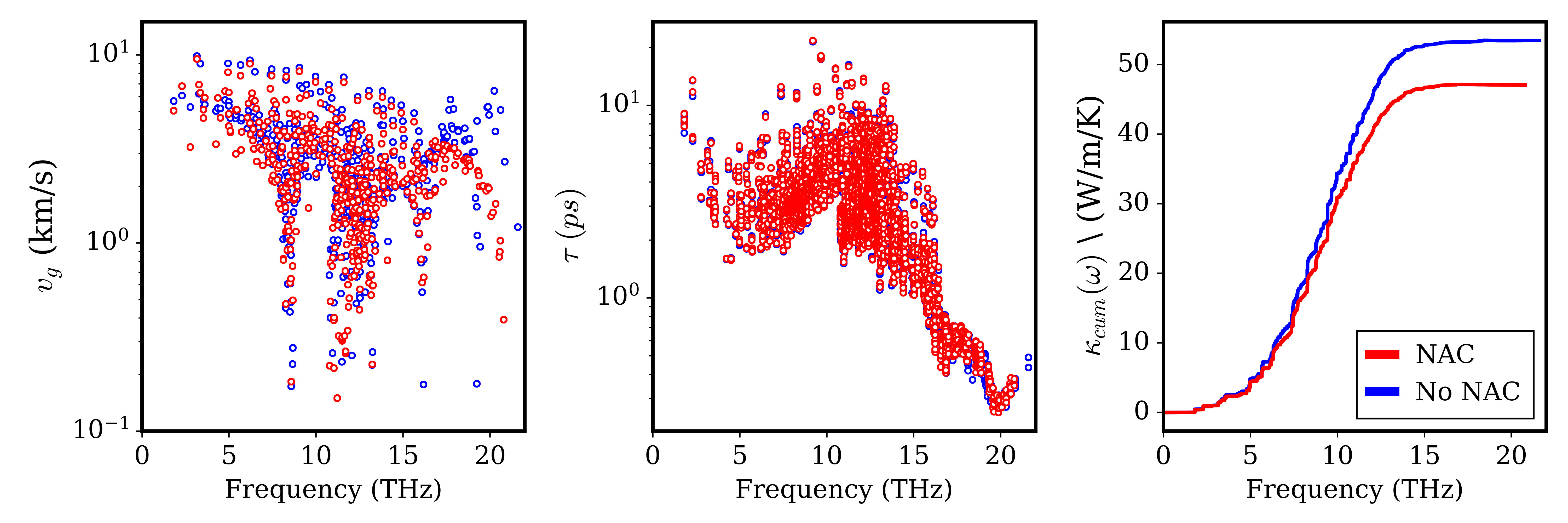}
    \caption{Phonon velocities (left), lifetimes (middle), and cumulative thermal conductivity (right) for \ce{MgO} with and without the NAC correction.}
    \label{fig:mgo_kc_lifetime_mfp}
\end{figure}

Lattice dynamics calculations were carried out on rock-salt \ce{MgO} at \SI{300}{\kelvin} to assess the role of long-range Coulomb interactions on the thermal transport of a polar material. The  second- and third-order force constants are obtained with DFPT\cite{giannozzi_advanced_2017, paulatto2013anharmonic} using the norm-conserving Vanderbilt pseudopotentials for Mg and O\cite{hamann2013optimized}. Figure \ref{fig:mgo_dispersion} shows the phonon frequency dispersion with and without the inclusion of non-analytical corrections (NAC)\cite{gonze1997dynamical}. Near $\Gamma$, the longitudinal-optical (LO) branch is hardened, opening a clear LO–TO split, while leaving the acoustic branches unchanged.\\
This shift leads to a small suppression in the thermal conductivity, primarily through the diminished contribution of optical phonons as seen in the plot of the cumulative $\kappa$ ($\kappa_{cum}$) against frequency in Figure \ref{fig:mgo_kc_lifetime_mfp}. The cumulative thermal conductivity $\kappa_{cum}$ exhibits a steeper initial rise and converges to a higher value without NAC ($\sim 5$ W/m/K higher), indicating that optical and mid-frequency acoustic modes are over-contributing when the LO-TO splitting is neglected. The reduction occurs primarily through reduced mode velocities, as the lifetimes are almost identical.

\begin{figure}[h!]
    \centering
    \includegraphics[width=0.50\textwidth]{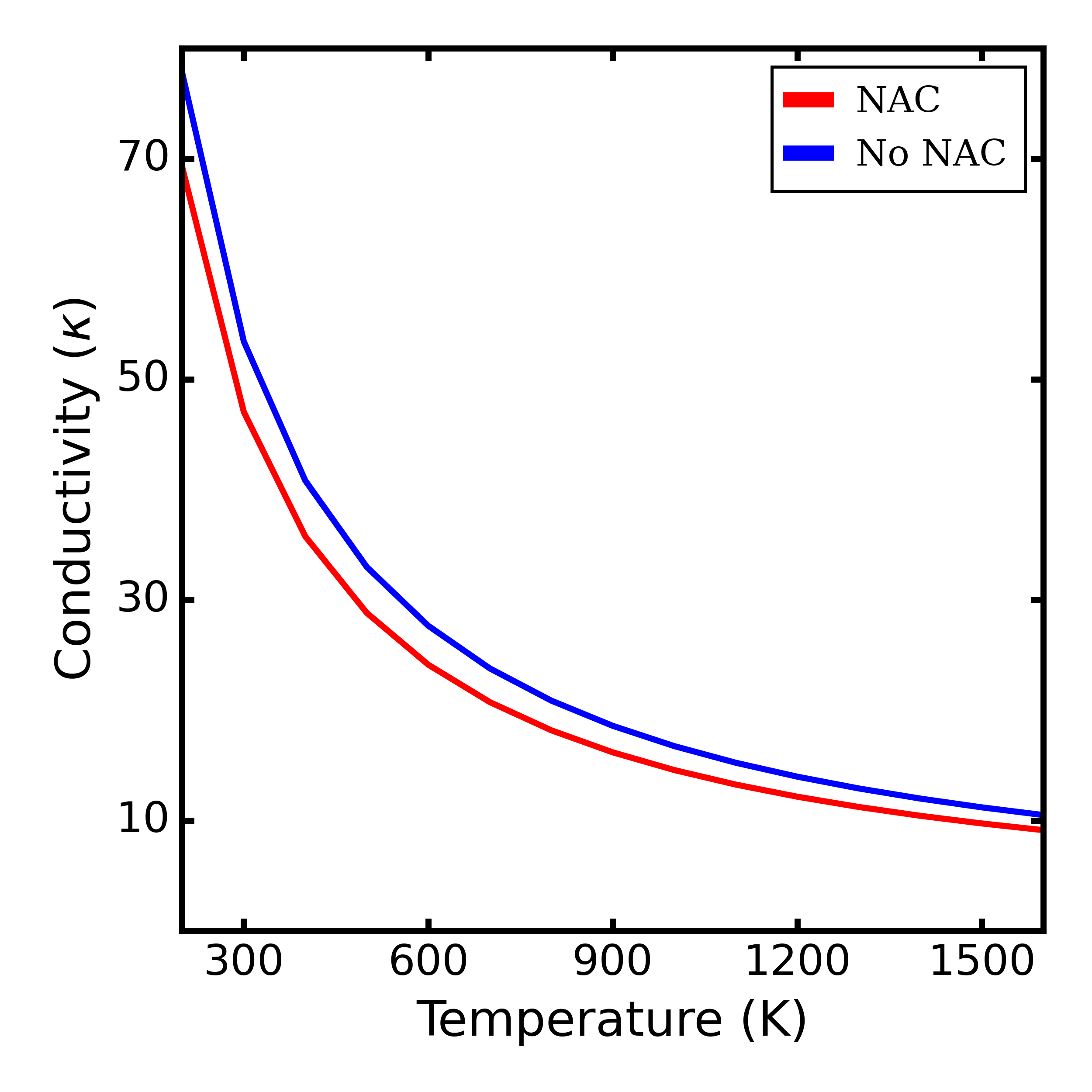}
    \caption{Thermal conductivity versus temperature calculated from full inversion of the scattering matrix for \ce{MgO} with and without the NAC correction.}
    \label{fig:mgo_kc_vs_T}
\end{figure}

The $\kappa$ of both systems was measured across a broad temperature span in Figure \ref{fig:mgo_kc_vs_T}.
Thermal conductivity falls proportionally with $T^{-1}$ for both calculations at low temperature, consistent with Umklapp-dominated transport. The NAC curve stays systematically lower than the result without corrections across the entire range.  The gap is largest at low temperature, where long-lived optical modes have the highest impact. The gap slowly narrows at high-T as strong phonon–phonon scattering washes out details of the spectrum.

\subsection{\revtwo{Thermal Transport in Amorphous Silica}}
\label{sec:amorphous-sio2}

\revtwo{All of our applications thus far have been crystalline systems with periodicity and symmetry. However, a central motivation for $\kappa$ALDo~2.0 is the treatment of materials that lack long-range order and whose phonon modes cannot be described by Bloch states, where the Boltzmann transport equation does not apply. Amorphous silica (\ce{SiO2}) is a representative test case for this regime. Its vibrational modes separate into propagons, which resemble propagating phonons at low frequency, diffusons, which cannot be associated to a well-defined wavevector and carry heat through coherent resonances, and locons, which are localized and carry little heat~\cite{allen1999diffusons}. The quasi-harmonic Green-Kubo solver of Section~\ref{sec:architecture} treats all three on the same footing through the generalized velocity operator introduced with Eq.~\ref{eq:qhgk}, and reduces to the Allen-Feldman result in the harmonic limit.}

\revtwo{We illustrate this on a 648-atom cubic cell of amorphous silica (216 \ce{Si} and 432 \ce{O}), with second- and third-order force constants obtained by finite differences using the Vashishta potential~\cite{vashishta_interaction_1990} through the LAMMPS interface. The cell is relaxed to its energy minimum using the FIRE optimization scheme, and the QHGK calculation is run at the $\Gamma$ point ($N_q = 1$, $N = 648$) at \SI{300}{\kelvin} with Gaussian broadening, following the setup of Fiorentino, Pegolo, and Baroni~\cite{fiorentino_hydrodynamic_2023}.}

\begin{figure}[H]
    \centering
    \begin{subfigure}{0.32\textwidth}
        \includegraphics[width=\linewidth]{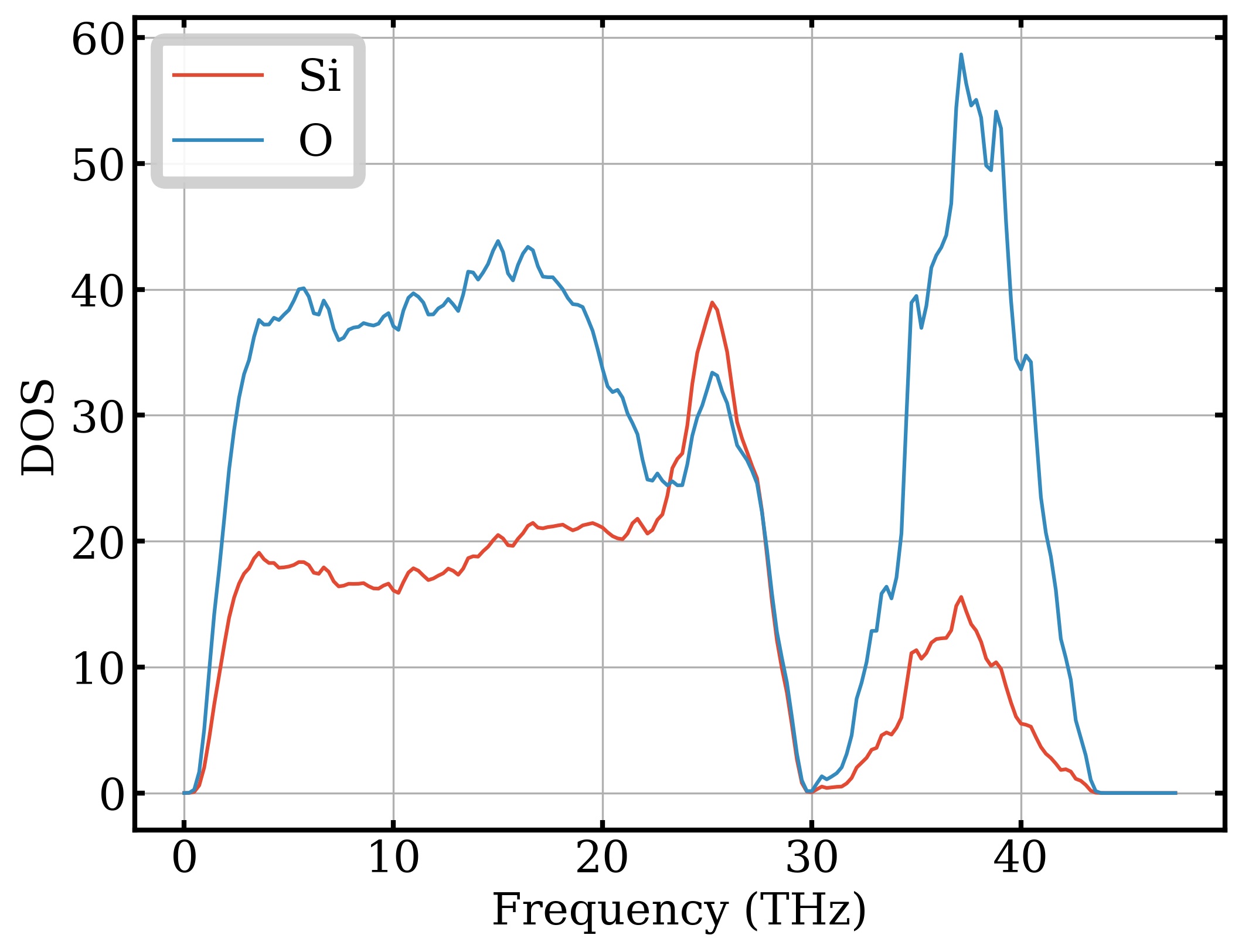}
        \caption{}
        \label{fig:asio2_dos}
    \end{subfigure}
    \hfill
    \begin{subfigure}{0.32\textwidth}
        \includegraphics[width=\linewidth]{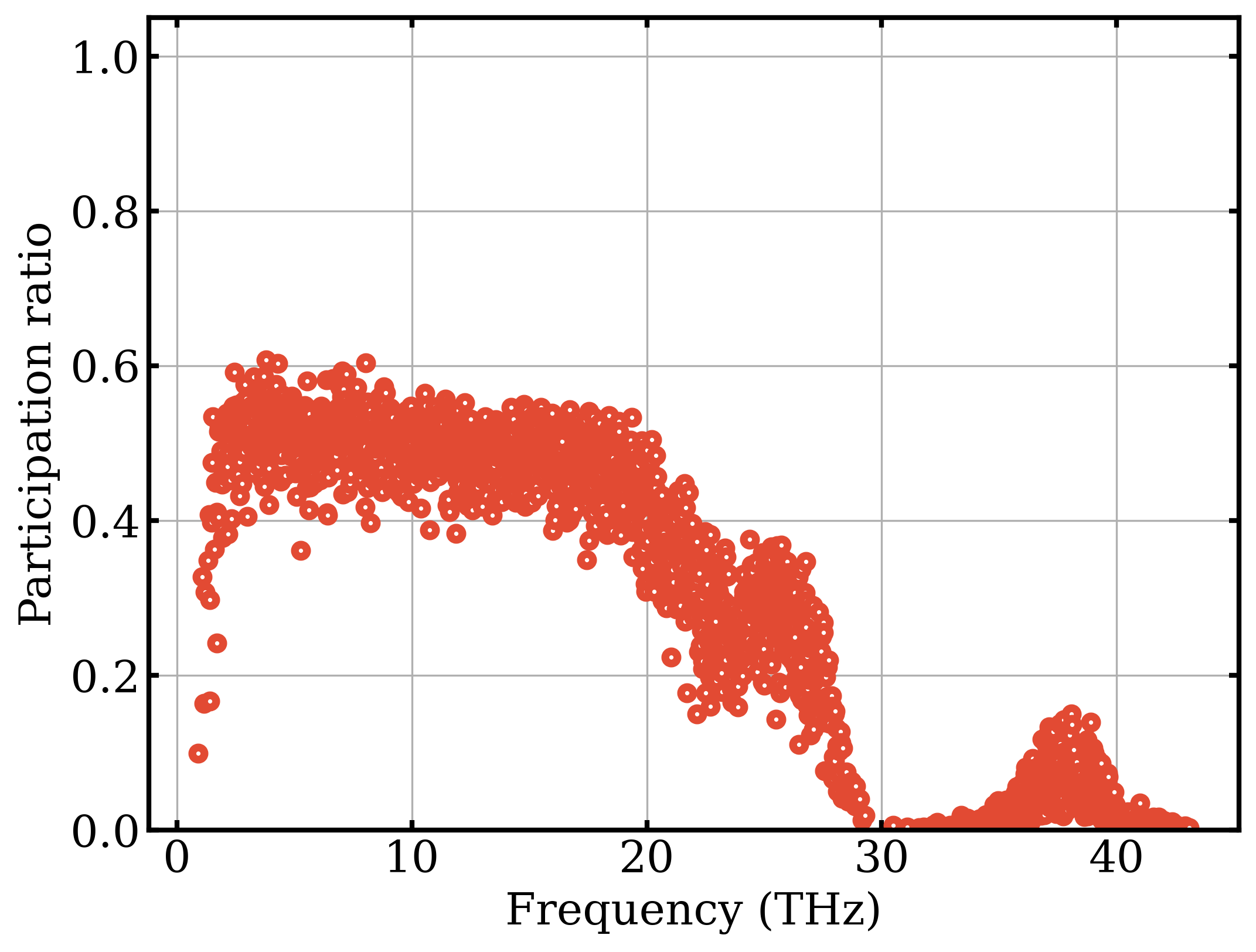}
        \caption{}
        \label{fig:asio2_pr}
    \end{subfigure}
    \hfill
    \begin{subfigure}{0.32\textwidth}
        \includegraphics[width=\linewidth]{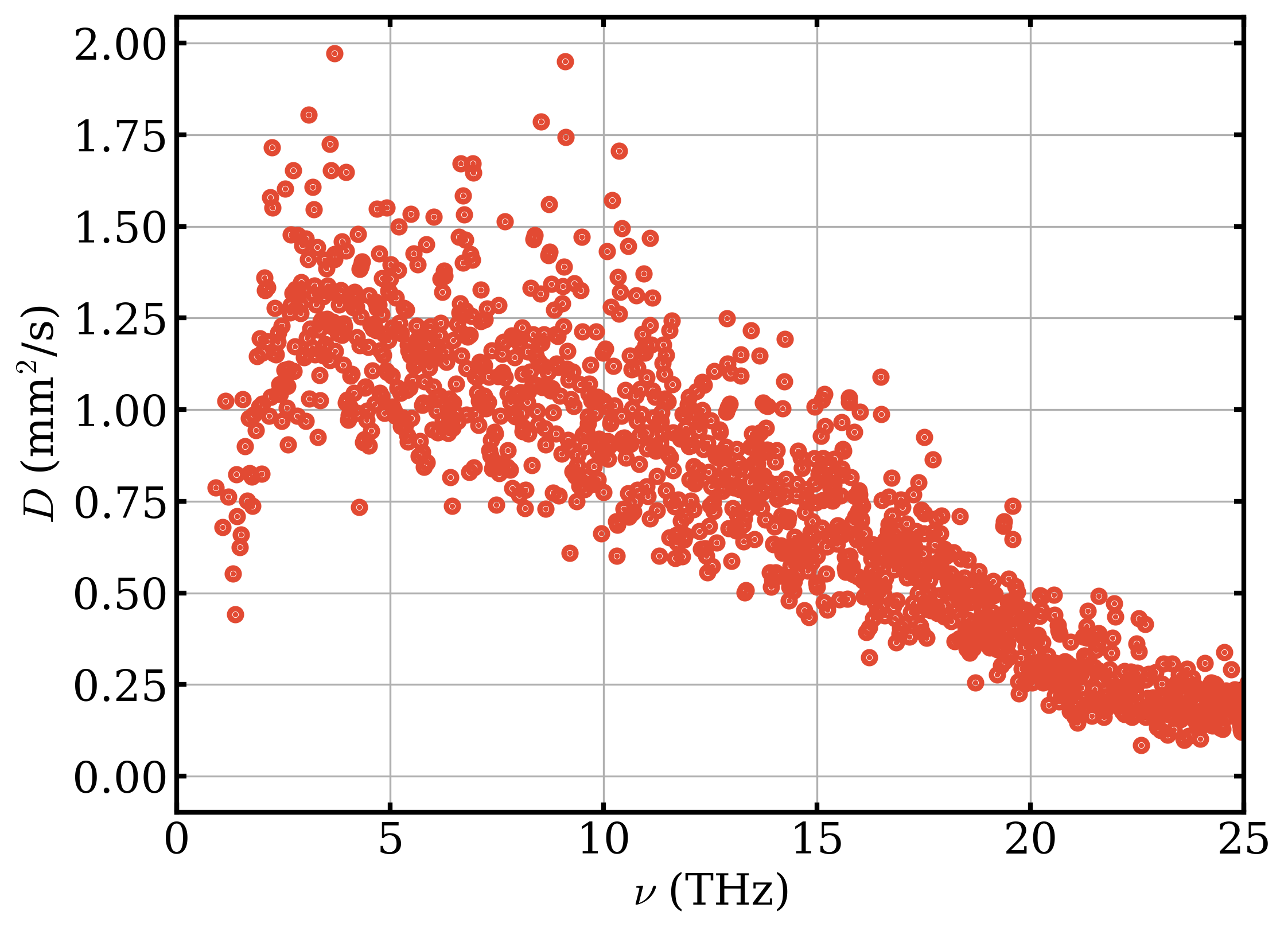}
        \caption{}
        \label{fig:asio2_diff}
    \end{subfigure}
    \caption{\revtwo{Vibrational character of amorphous silica (\ce{SiO2}), 648-atom cell at \SI{300}{\kelvin}. (a) Species-projected density of states for \ce{Si} and \ce{O}. (b) Participation ratio versus frequency, separating extended modes ($PR \sim 0.5$) from localized locons above \SI{30}{\tera\hertz}. (c) Mode-resolved thermal diffusivity versus frequency.}}
    \label{fig:amorphous_sio2_modes}
\end{figure}

\revtwo{Figure~\ref{fig:amorphous_sio2_modes} characterizes the vibrational modes. The species-projected density of states (a) shows the two bands typical of silica: a broad band below \SI{30}{\tera\hertz} with mixed \ce{Si} and \ce{O} character, and a high-frequency band near \SI{37}{\tera\hertz} dominated by oxygen, corresponding to the \ce{Si}-\ce{O} bond-stretching modes, separated by a gap around \SI{30}{\tera\hertz}. The participation ratio (b) stays near $0.5$ throughout the lower band, marking spatially extended propagons and diffusons, then collapses toward zero above \SI{30}{\tera\hertz}, identifying the high-frequency band as localized locons. The mode diffusivity (c) is largest for the low-frequency modes and decreases monotonically with frequency, consistent with diffuson-dominated transport.}

\begin{figure}[H]
    \centering
    \begin{subfigure}{0.32\textwidth}
        \includegraphics[width=\linewidth]{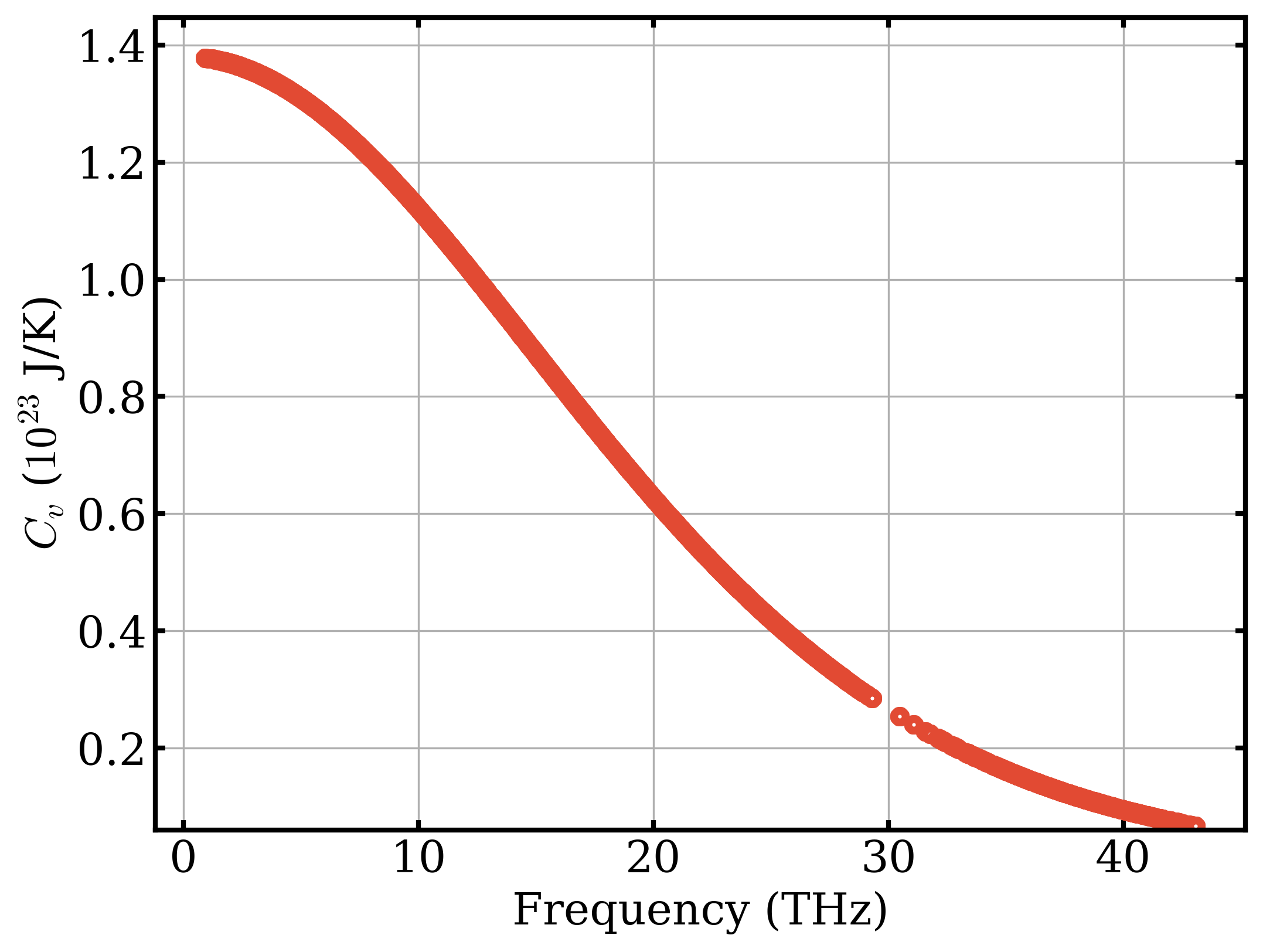}
        \caption{}
        \label{fig:asio2_cv}
    \end{subfigure}
    \hfill
    \begin{subfigure}{0.32\textwidth}
        \includegraphics[width=\linewidth]{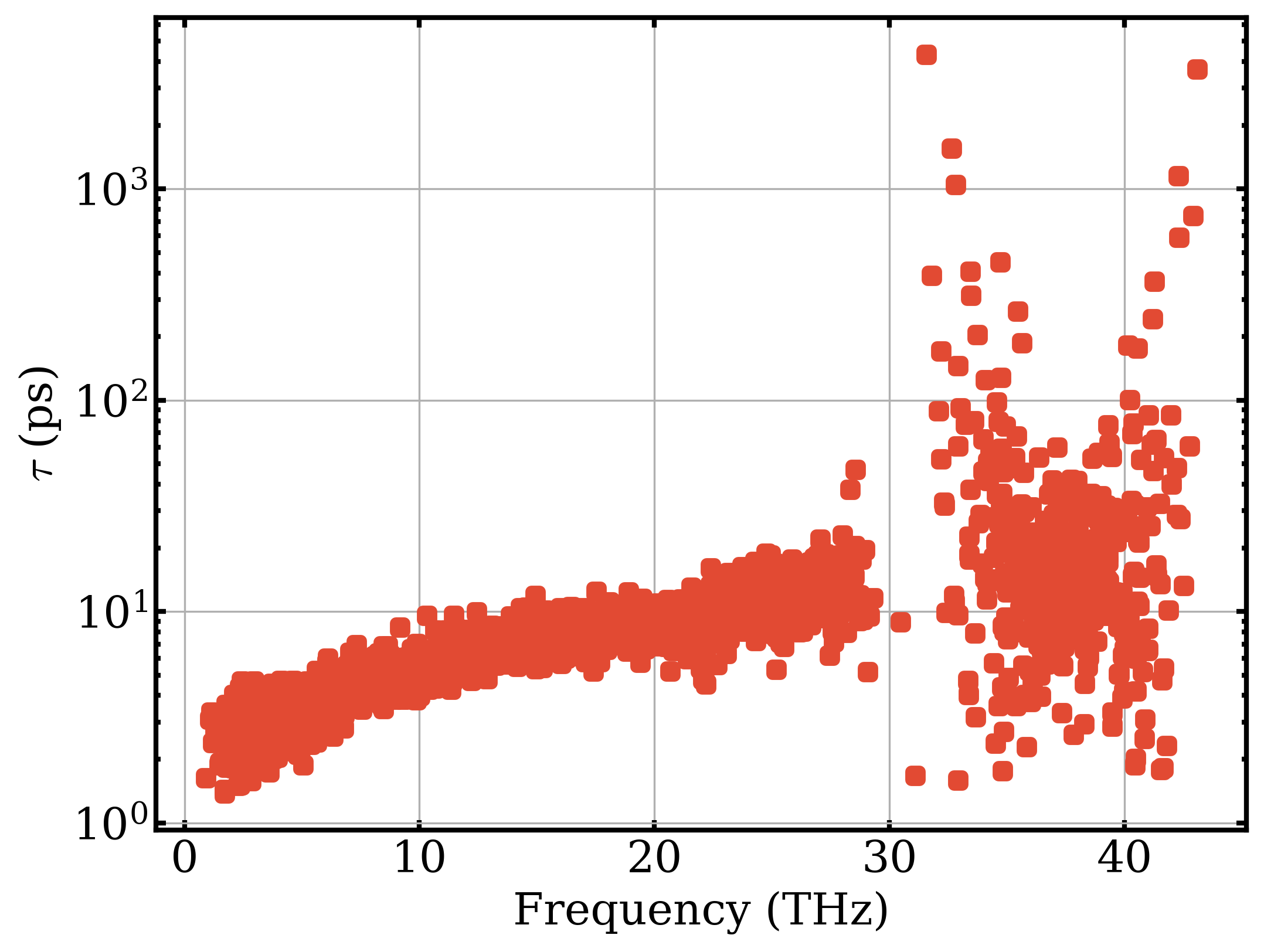}
        \caption{}
        \label{fig:asio2_lifetime}
    \end{subfigure}
    \hfill
    \begin{subfigure}{0.32\textwidth}
        \includegraphics[width=\linewidth]{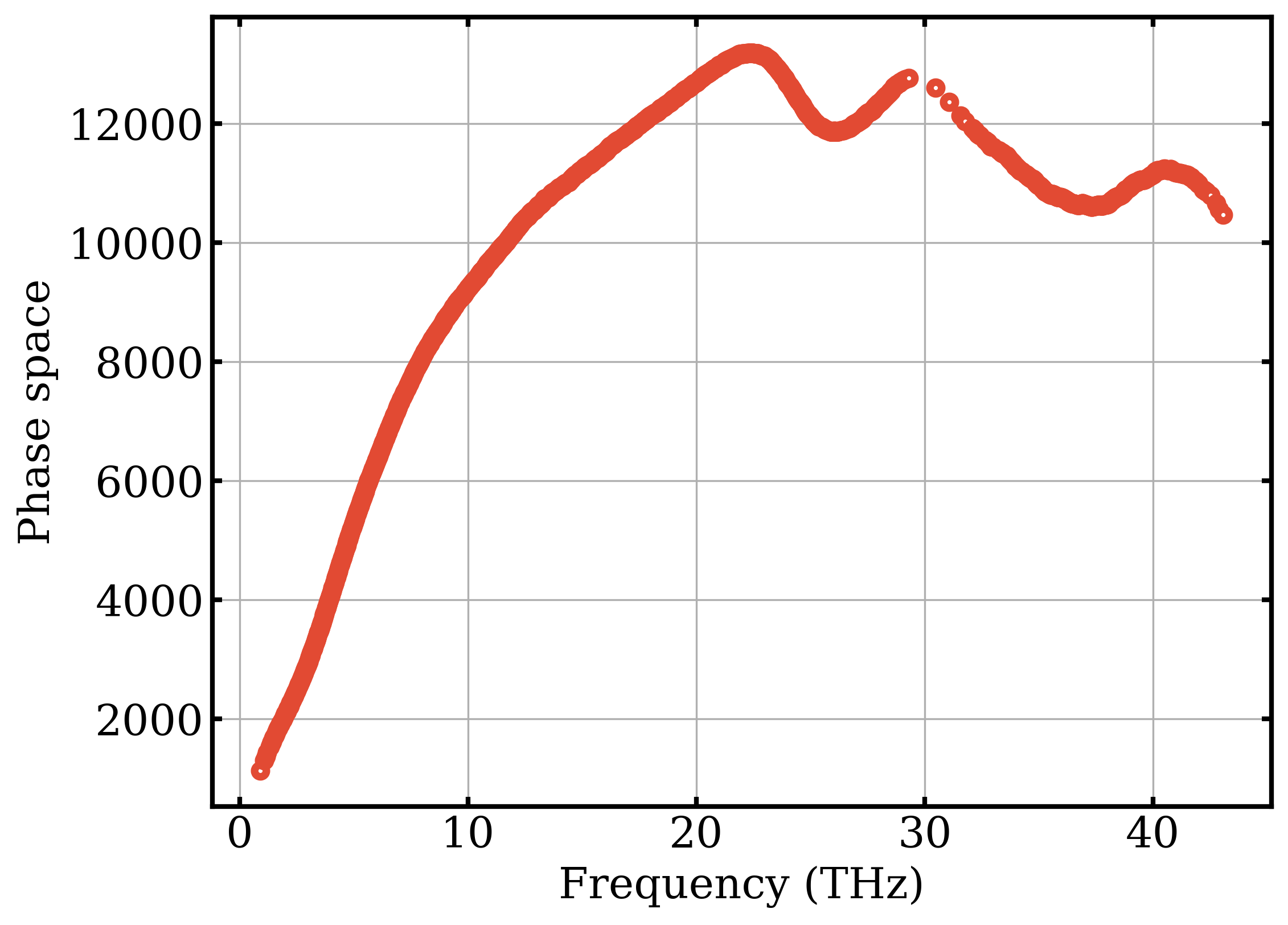}
        \caption{}
        \label{fig:asio2_ps}
    \end{subfigure}
    \caption{\revtwo{Per-mode properties of amorphous silica entering the QHGK conductivity. (a) Modal heat capacity versus frequency. (b) Mode lifetimes versus frequency. (c) Three-phonon scattering phase space versus frequency.}}
    \label{fig:amorphous_sio2_props}
\end{figure}

\revtwo{Figure~\ref{fig:amorphous_sio2_props} reports the per-mode quantities that enter the QHGK conductivity. The modal heat capacity (a) follows the quantum Bose-Einstein occupation, falling smoothly from its classical value at low frequency as the high-frequency modes freeze out at \SI{300}{\kelvin}. The mode lifetimes (b) grow from a few picoseconds at low frequency to tens of picoseconds, with the isolated locon band showing a broad scatter of long lifetimes. The three-phonon scattering phase space (c) rises steeply from the acoustic region and saturates across the diffuson band.}

\begin{figure}[H]
    \centering
    \begin{subfigure}{0.48\textwidth}
        \includegraphics[width=\linewidth]{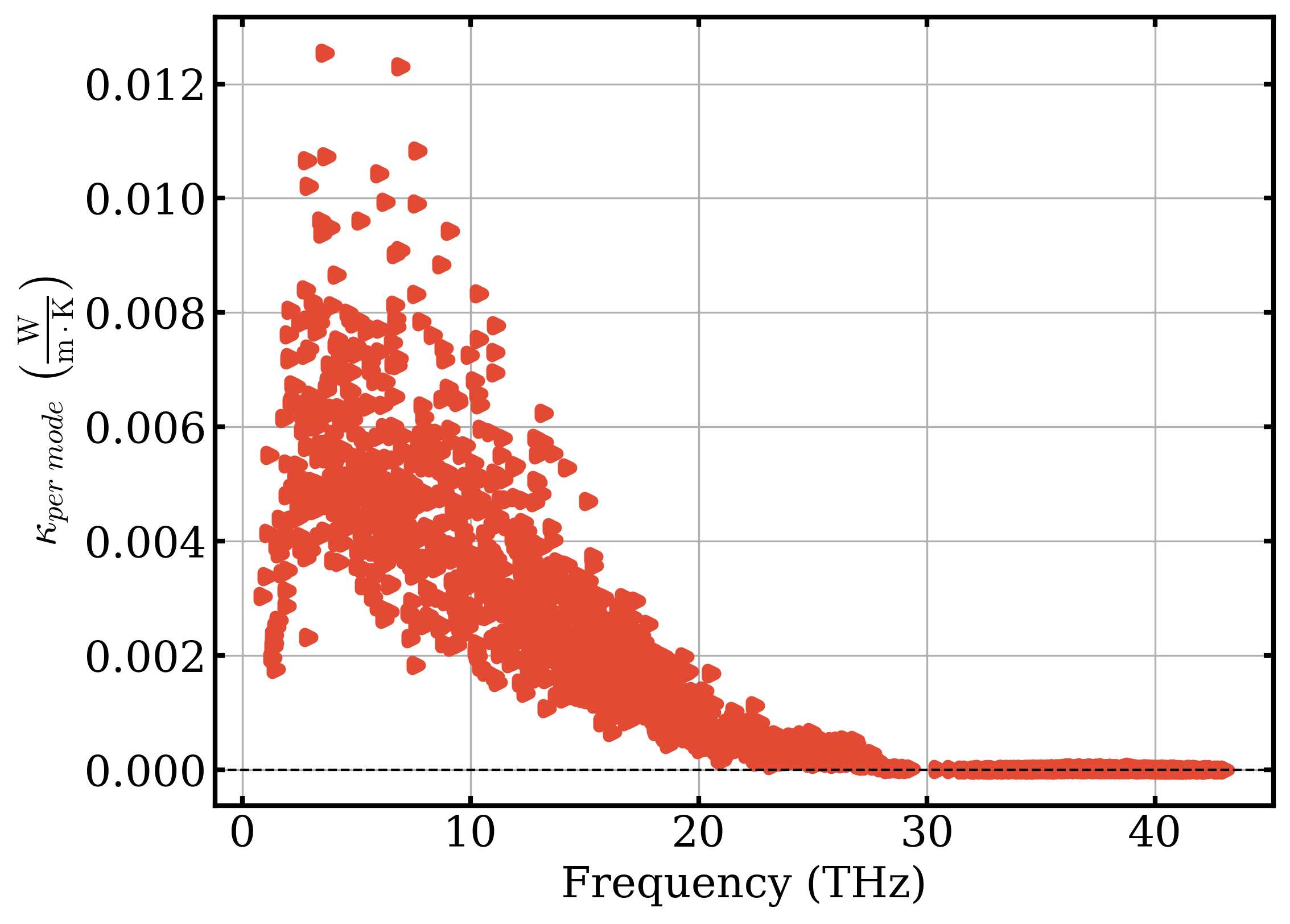}
        \caption{}
        \label{fig:asio2_kpm}
    \end{subfigure}
    \hfill
    \begin{subfigure}{0.48\textwidth}
        \includegraphics[width=\linewidth]{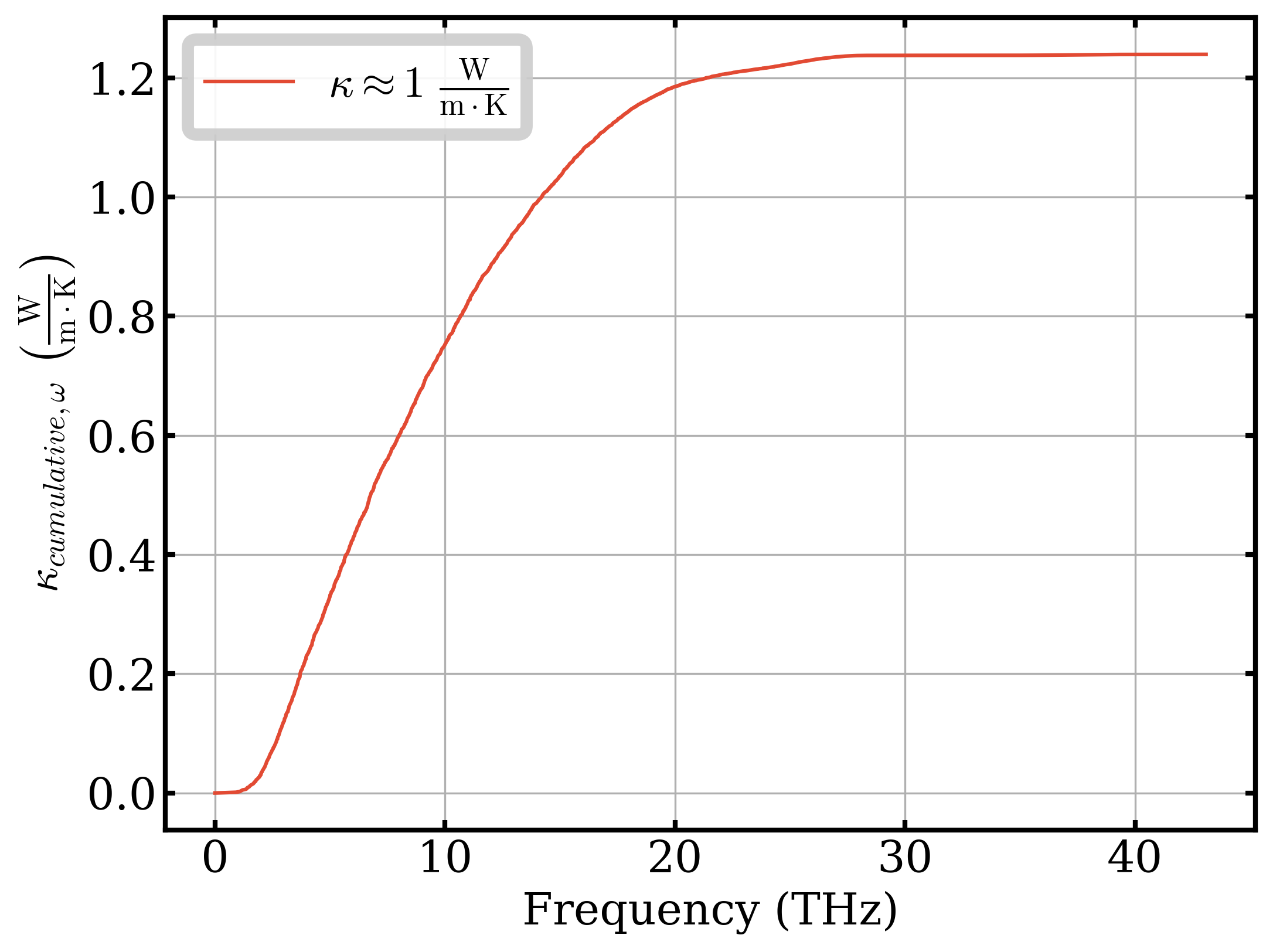}
        \caption{}
        \label{fig:asio2_kcum}
    \end{subfigure}
    \caption{\revtwo{Thermal conductivity of amorphous silica (\ce{SiO2}), 648-atom cell at \SI{300}{\kelvin}. (a) Per-mode contribution to $\kappa$ versus frequency. (b) Cumulative $\kappa$ versus frequency, saturating at $\kappa \approx \SI{1.15}{\watt\per\meter\per\kelvin}$.}}
    \label{fig:amorphous_sio2_kappa}
\end{figure}

\revtwo{The resulting thermal conductivity is $\kappa \approx \SI{1.15}{\watt\per\meter\per\kelvin}$, with a nearly isotropic tensor, in good agreement with the experimental value of fused silica ($\approx$ \SIrange{1.3}{1.4}{\watt\per\meter\per\kelvin}) and with the slight underestimate expected from the Vashishta potential. Figure~\ref{fig:amorphous_sio2_kappa} shows how this value is assembled: the per-mode contributions (a) are concentrated below \SI{15}{\tera\hertz}, and the cumulative conductivity (b) rises through the propagon and diffuson region and saturates by about \SI{28}{\tera\hertz}, with the localized high-frequency modes contributing negligibly. This confirms that, in amorphous silica, heat is carried almost entirely by the extended low-frequency vibrations, the regime where the QHGK treatment of diffuson transport is essential.}

\section{Conclusions and Outlook}
\label{sec:conclusion}

We have presented $\kappa$ALDo v2.0, a comprehensive open-source platform for thermal transport calculations from anharmonic lattice dynamics. The software provides researchers with a unified framework combining (i) flexible force constant import from DFT, empirical potentials, and machine-learned potentials, (ii) efficient implementations of BTE (RTA, self-consistent, full inversion, eigendecomposition) and QHGK methods, (iii) essential physical corrections (isotopic scattering, non-analytical terms), (iv) GPU acceleration and sparse tensor operations for large systems, and (v) modular Python design with extensive documentation and deployment tools.

Version 2.0 introduces significant advances over the original release~\cite{barbalinardo2020efficient}, including native TDEP support for temperature-dependent force constants, seamless MLP integration via ASE and LAMMPS, flexible storage backends enabling 10,000+ atom simulations, anharmonicity diagnostics ($\sigma_A$ score), and comprehensive testing/continuous integration. These enhancements position $\kappa$ALDo~2.0 as a premier tool for thermal transport research in complex materials.

Applications to \ce{CsPbBr3} and \ce{MgO} demonstrate the software's capabilities for challenging systems: perovskites requiring finite-temperature stabilization via TDEP and polar oxides needing non-analytical corrections. These examples illustrate workflows typical of contemporary materials science, combining multiple simulation packages and validation against experimental data.

Future development directions include: (i) hybrid MPI+GPU parallelization for multi-node scaling, (ii) automation of convergence testing for $q$-meshes and broadening parameters, (iii) four-phonon scattering for materials where three-phonon processes are insufficient (e.g., Si, diamond)~\cite{feng2017four}, (iv) coupling to excited-state dynamics for electron-phonon and exciton-phonon interactions, (v) integration with high-throughput frameworks (AiiDA, Atomate) for database generation\revtwo{, and (vi) optional reduced- and mixed-precision execution as an opt-in for users who prioritize throughput over numerical precision}.

We invite the community to contribute code, report issues, and propose new features via the GitHub repository. A discussions page (\url{https://github.com/nanotheorygroup/kaldo/discussions}) is available for questions, feature requests, and sharing workflows. With its open-source license, comprehensive documentation, and modular architecture, $\kappa$ALDo~2.0 aims to serve both as a production tool for applied research and a platform for method development in phonon transport theory.

\section*{Data Availability}

Input files, Jupyter notebooks, and results for all examples presented in this paper are available in the examples repository. A Docker image containing $\kappa$ALDo 2.0 and all dependencies can be obtained via \texttt{docker pull gbarbalinardo/kaldo:latest}.

\section*{CPC Program Library Compliance}

This manuscript adheres to the requirements specified in the Computer Physics Communications Program Library guidelines. The submitted package includes: (i) source code with test suite, (ii) installation instructions, (iii) example input/output files, (iv) auto-generated documentation, and (v) containerized environment for reproducibility.

\section*{Acknowledgments}


We are grateful to Riccardo Dettori, Frank Cerasoli, and Nicholas Martinez for helpful discussions and beta testing, and Mattias Perez and Higo de Araujo Oliveira for assistance in refining the example repository. 
This research was supported by the U.S. Department of Energy, Office of Basic Energy Sciences, Division of Materials Science and Engineering, grant DE-SC0022288.
MIT Office of Research Computing and Data provided computing resources for performance benchmarks.

\bibliography{bibliography}

\end{document}